\documentclass[aip,groupedaddress,reprint]{revtex4-1}

\usepackage{color}

\usepackage{graphicx}
\usepackage{dcolumn}
\usepackage{bm}
\usepackage{hyperref}

\usepackage{fancyhdr}
\begin{document}

\title{Combined hybrid functional and DFT+$U$ calculations for metal chalcogenides}
\author{Mehmet Aras}
\affiliation{
Department of Physics, Gebze Institute of Technology, Gebze, Kocaeli 41400, Turkey
}
\author{\c{C}etin K{\i}l{\i}\c{c}}
\email[E-mail: ]{cetin\_kilic@gyte.edu.tr}
\affiliation{
Department of Physics, Gebze Institute of Technology, Gebze, Kocaeli 41400, Turkey
}

\date{
6 July 2014
}

\begin{abstract}
In the density-functional studies of materials
  with {\it localized} electronic states, 
  the local/semilocal exchange-correlation functionals are often
  either combined with a Hubbard parameter $U$ as in the LDA+$U$  method 
  or mixed with a fraction of exactly computed (Fock) exchange energy yielding a hybrid functional.
Although some inaccuracies of the semilocal density approximations are thus fixed to a certain extent,
  the improvements are not sufficient to make the predictions agree with the experimental data.
Here we put forward the perspective that the hybrid functional scheme and the LDA+$U$ method should be treated as {\it complementary}, and
  propose to combine the range-separated (HSE) hybrid functional with the Hubbard $U$.
We thus present a variety of HSE+$U$ calculations for a set of II-VI semiconductors,
  consisting of zinc and cadmium monochalcogenides,
  along with comparison to the experimental data.
Our findings imply that an {\it optimal} value $U^\ast$ of the Hubbard parameter
  could be determined, which ensures that the HSE+$U^\ast$ calculation reproduces the experimental band gap.
It is shown that
  an improved description
  not only of the electronic structure
  but also of the crystal structure and energetics
  is obtained by adding the $U^\ast$ term to the HSE functional,
  proving
  the utility of HSE+$U^\ast$ approach
  in modeling semiconductors with localized electronic states.

\end{abstract}

\maketitle

\thispagestyle{plain}

%
%
\section{\label{giris}Introduction}

Static correlation,\cite{handy01,becke13}
  arising from the tendency of electrons to distribute themselves over the various centers,
  is pronounced in materials containing localized {\it d} or {\it f} electrons
  such as some transition-metal or rare-earth compounds.
The local density approximation\cite{kohn65} (LDA) or the generalized gradient approximation\cite{PBE} (GGA)
  commonly employed in Kohn-Sham density functional theory\cite{kohn65} (DFT)
  inherently assume a {\it localized} exchange-correlation hole,
  implying that static correlation is treated in an unrestrained manner in these approximations.\cite{becke05}
Thus local or semilocal exchange-correlation energy $E_{\rm xc}$ functionals are often
  either combined with a Hubbard parameter $U$ as in the LDA+$U$ method\cite{anisimov97}
  or mixed with a fraction $\alpha$ of exactly computed \cite{becke93} (Fock) exchange energy $E_{\rm x}^{\rm exact}$,
  yielding a hybrid functional
  \begin{equation}
    E_{\rm xc}^{\rm hybrid}=E_{\rm x}^{\rm exact}+(1-\alpha)(E_{\rm x}^{\rm GGA}-E_{\rm x}^{\rm exact})+E_{\rm c}^{\rm GGA},
  \end{equation}
  where the second term models the static correlation energy.\cite{csonka10}
For $\alpha > 0$, the static correlation energy is reduced
  in favor of the suppression of electron fluctuations,
  leading to a better description for the localized electron states
 (as evidenced by the improved prediction of
  the binding energy of localized $d$ states,\cite{oba08,wrobel09,burbano11,pozun11,li12,zhang12}
                                   band gaps,\cite{heyd05,marsman08,oba08,wrobel09,burbano11,wu11,lucero12,li12,zhang12,pozun11,kanan12}
                         and magnetic moments\cite{rodl09,liao11,chen12,heinemann13,pozun11,kanan12}).
In the LDA+$U$ approach, where a $d$ ion is treated as an open system with fluctuation number of electrons,\cite{anisimov97}
  a term including $U$ is added to the total energy,
  which penalizes more fluctuating configurations\cite{perdew07,perdew09}
  and therefore leads to
  a better description of the localized states
 (as evidenced by the improved prediction of
  the binding energy of localized $d$ states,\cite{rohrbach03,mikaye06,petersen06,zhang12,zhang13}
                                   band gaps,\cite{anisimov97,dudarev98,rohrbach03,rohrbach04,mikaye06,gupta07,devey09,rodl09,liao11,arroyo11,zhang12,zhang13,kanan12}
                         and magnetic moments\cite{anisimov97,rohrbach03,rohrbach04,rollmann04,wang06,petersen06,gupta07,devey09, liao11,zhang13,kanan12}).
Thus the hybrid functional scheme and the LDA+$U$ approach could be regarded
  as alternative means\cite{tran06,jollet09} for fixing inaccuracies of the semilocal density approximations,
  which result from insufficient localization of $d$ electrons.
Indeed, it has recently been proposed\cite{hong12,andriotis13} to derive the value of $U$ from hybrid functional calculations.
In contrast, we think it is appropriate to adopt
  a perspective where the hybrid-functional and DFT+$U$ methods are treated {\it complementary}
  (inasmuch as they both reduce the static correlation energy),
  which led us to combine hybrid functionals with the Hubbard $U$.
From a different point of view, this means that
  one of the two methods (DFT+$U$) is utilized
  to reduce the {\it residual} self-interaction error\cite{perdew81}
  of the other one (hybrid-functional),
  which is pragmatically justified.
Furthermore,
  Iv\'ady {\it et al.} (Ref.~\onlinecite{ivady14}) have recently shown that
  a hybrid exchange-correlation potential
  could be cast into a mathematical form
  that is reminiscent of the on-site Hubbard potential
  for a subsystem of {\it localized} orbitals,
  providing theoretical justification for our methodology:
An additional on-site (DFT+$U$) potential is added
  to the hybrid exchange-correlation potential,
  which is applied {\it only} to strictly localized states.
This improves the physical description
  because localized $d$-band states and delocalized crystal states are differentiated in the hybrid-functional+$U$ approach,
  which are indifferent to the hybrid functional itself.
It is also interesting in this regard to point out that
  the DFT+$U$ and hybrid-functional methods could both be regarded as approximations to the $GW$ method,\cite{hedin65}
  as articulated in Refs.~\onlinecite{anisimov97} and \onlinecite{henderson11}, respectively.
The incentive of using these two methods together
  is then to increase the level of approximation,
  provided that they are complementary.

\begin{figure}
  \begin{center}
    \resizebox{0.50\textwidth}{!}{%
      \includegraphics{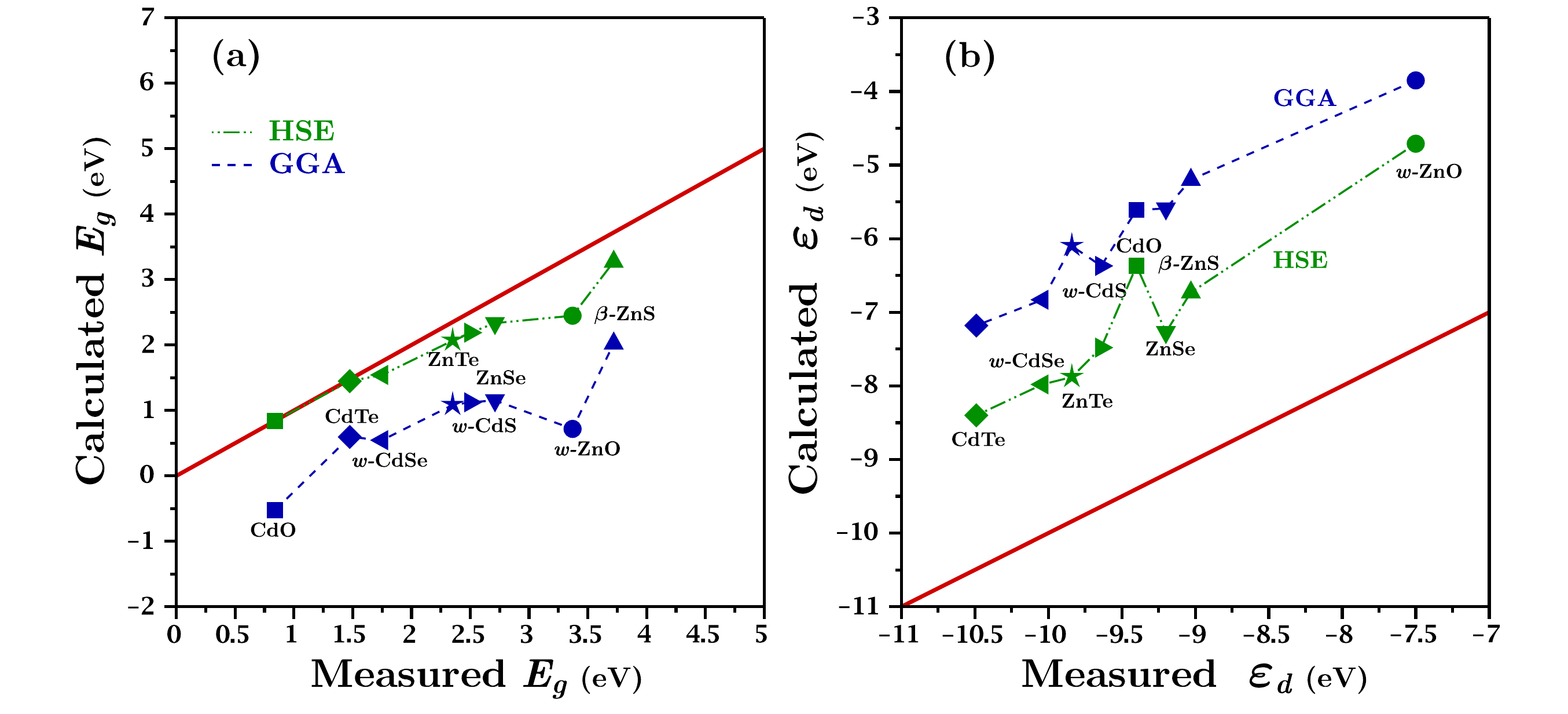}
    }
  \end{center}
  \vspace{-0.4cm}
  \caption{
          (Color online)
          Calculated versus measured values of  the band gap $E_g$ (a) and
                                            the $d$ band position $\varepsilon_d$ relative to the valence band maximum (b)
          for zinc and cadmium monochalcogenides.
          The experimental values of $E_g$ and $\varepsilon_d$
            are taken from Refs.~\onlinecite{bandgap1,bandgap2,bandgap3} and
                           Refs.~\onlinecite{dstate1,dstate2}, respectively.
          The values obtained from the present GGA (HSE) calculations are connected by blue dashed (red dot-dashed) lines to guide the eye.
          The solid black lines passes through the experimental values.
          }
   \label{Egcalvsexp}
\end{figure}

It is usually necessary to perform a calibration\cite{lutfalla11,krcha13}
  for the value of $U$ that is {\it optimal} with respect to the material properties under consideration.
Besides $U$ is not only element-specific\cite{lutfalla11} but also material-specific.\cite{karazhanov06,barcaro10}
Thus it is appealing to employ a hybrid functional
  with an exchange mixing coefficient $\alpha$
  that is in practice fixed to a single {\it universal} value,
  e.g., $\alpha=1/4$ in both global\cite{perdew96} and range-separated Heyd-Scuseria-Ernzerhof\cite{heyd03} (HSE) hybrid functionals.
It should, however, be noted that
  setting the optimal value for $\alpha$ as $1/4$ in Ref.~\onlinecite{perdew96} was accomplished {\it empirically}
  (via error analysis of the atomization energies),
  which would not necessarily be optimal for {\it other} material properties.\cite{vydrov06,park11,lim12,himmetoglu13}
We found,
  in line with earlier reports,\cite{paier06,fuchs07,henderson11}
  that the hybrid (HSE) functional calculations with $\alpha=1/4$ improve
  the prediction of both the $d$ band position $\varepsilon_d$ relative to the valence band maximum and the band gap $E_g$
  but these improvements are not sufficient to make the predictions agree with the experimental data.
This is demonstrated in Fig.~\ref{Egcalvsexp} for zinc and cadmium monochalcogenides,
 where the calculated and measured values of $E_g$ (left panel) and $\varepsilon_d$ (right panel) are plotted with respect to each other.
Figure~\ref{Egcalvsexp}(a) shows
  that
  (i) the improvement for the band gap is impressive
  for systems with a somewhat small band gap,
  and 
  (ii) the band gap is {\it still} significantly underestimated for wide band gap semiconductors such as ZnO.
As explored in Appendix,
  both the GGA band gap error $\Delta E_g^{\rm GGA}$ and the HSE correction $E_g^{\rm HSE}-E_g^{\rm GGA}$
  are inversely proportional to the high-frequency dielectric constant $\epsilon_\infty$
  so that $ \Delta E_g^{\rm GGA} \simeq A /\epsilon_\infty $ and 
          $E_g^{\rm HSE}-E_g^{\rm GGA} \simeq A^\prime /\epsilon_\infty $,
          where the constants $A$ and $A^\prime $ satisfy $A^\prime < A$.
Owing to the latter, the HSE improvement falls short for materials with relatively small dielectric constant
  (with the exception of CdO for which the HSE calculation yields the right direct and indirect band gaps, cf. Ref.~\onlinecite{burbano11}).
Figure~\ref{Egcalvsexp}(b) shows that
   the HSE-calculated $\varepsilon_d$ is {\it still} too high
   although there is a significant correction of about $1.3\pm0.4$~eV.
It should be noted that
  the prediction of $\varepsilon_d$
  could further be improved by adding a Hubbard $U$ term to the hybrid functional,
  which would enable one to adjust the $d$ band position.
It is also interesting to note that
  the measured values of $\varepsilon_d$ could indeed be reproduced
  by using adjusted $U$ values, cf. Fig. 3 of Ref.~\onlinecite{karazhanov06},
  in the case of zinc monochalcogenides.
These observations also motivate us to treat
  the hybrid functional scheme and the DFT+$U$ method
  as {\it complementary} rather than {\it alternative} approaches.
Accordingly,
  we propose here to combine the screened hybrid functional of Heyd, Scuseria, and Ernzerhof with the Hubbard $U$.
The main advantage of the latter is that
  strictly localized and delocalized states are {\it screened differently}
  since {\it only} the former are subject to an additional on-site (DFT+$U$) potential.\cite{ivady14,ivady13}
In contrast, localized and delocalized states are
  indifferent to the original HSE functional
  as long as the same set of parameters, viz. the exchange mixing coefficient $\alpha$ and the screening parameter $\omega$, are used
  for {\it all} states.
Additionally,
  we regard $U$ as a semiempirical parameter,
  in line with the perspective\cite{albers09} that
  the Hubbard term added to the density functionals
  is essentially a phenomenological many-body correction.
Our findings show that
  the HSE+$U$ calculations
  performed by using an {\it adjusted} $U$ value
  reproduce the measured band gap and, at the same time,
  result in an improved physical description
  not only of the electronic structure
  but also of the crystal structure and energetics
  for the semiconductors with localized $d$ electrons.
This is obviously very convenient
  for practical purposes such as setting the range of the electron chemical potential accurately
  in the point defect calculations, e.g., Ref.~\onlinecite{oba11}.
It is also very convenient
   because it enables one to employ the measured $E_g$, instead of $\varepsilon_d$, in setting the $U$ value.
Note that there is usually some scatter in the measured data for $\varepsilon_d$,
   which partly reflects the fact that the width of the $d$ bands is nonzero no matter how localized the states are.

The underestimation of the band gap in the HSE calculations, cf. Fig.~\ref{Egcalvsexp}(a),
  could partially be attributed to lacking the correlation part of the discontinuity of the exchange-correlation potential.\cite{seidl96}
Similarly,
  the discontinuity of the exchange-correlation potential
  is not fully restored in the LDA/GGA+$U$ calculations
  even though the $U$ term added to the density functionals yields a discontinuous contribution.\cite{anisimov93}
It should also be commented that
  setting the right value of $U$ {\it empirically} is not straightforward
  because one needs
  to take accounts of hybridization and screening of $d$ electrons {\it a priori}.
Furthermore,
  the measured value of $E_g$ could not be reproduced no matter how large a value of $U$ is used
  in the LDA+$U$ calculations performed for zinc monochalcogenides,
  cf. Fig. 3 of Ref.~\onlinecite{karazhanov06}.
Our study provides a
  resolution to this difficulty with the aid of hybrid functional,
  and proves that an adequate $U$ value could be determined
  by simply matching the experimental band gap.

The rest of the paper is organized as follows:
The next section is devoted to the method of calculation,
  which also summarizes the computational details.
This is followed by a discussion of the calculation results
  before concluding remarks given in the last section.

%
%
\section{\label{yontem}Method}

All calculated properties reported here were obtained
  via semilocal or hybrid DFT calculations
  using the Perdew-Burke-Ernzerhof\cite{PBE} (PBE) or Heyd-Scuseria-Ernzerhof\cite{heyd03} (HSE) functionals, respectively.
In the hybrid functional calculations,
  we employed the HSE06\cite{HSE06} functional by setting the screening parameter\cite{HSE06,wrobel09} $\omega=0.207$ \AA$^{-1}$
  (and exchange mixing coefficient $\alpha=0.25$ as implied in Section~\ref{giris}).
In the HSE+$U$ calculations we used the simplified (rotationally invariant) approach\cite{dudarev98}
  where the difference between the on-site Coulomb $\bar{U}$ and exchange $\bar{J}$ parameters
  is employed as the {\it effective} Hubbard parameter $U=\bar{U}-\bar{J}$.
We performed a variety of calculations for zinc and cadmium monochalcogenides
  by employing the projector augmented-wave (PAW) method,\cite{PAW}
  as implemented in VASP code.\cite{VASP,VASPpaw}
The 2s and 2p,
    3s and 3p,
    4s and 4p,
    5s and 5p,
    3d and 4s, and
    4d and 5s states are treated
  as valence states for 
    oxygen,
    sulfur,
    selenium,
    tellurium,
    zinc, and 
    cadmium, respectively.
Plane wave basis sets were used to represent the electronic states,
  which were determined 
  by imposing a kinetic energy cutoff of 520 eV for the systems that include oxygen atoms and
                                         400 eV for the rest of the systems.

We first carried out optimization of the crystal structures
  where concurrent relaxations of the cell volume and shape as well as the ionic positions
  were performed until the total energy was converged within 1 meV
  and the maximum value of residual forces on atoms was reduced to be smaller than 0.01 eV/\AA.
In these optimizations,
  we used the primitive unit cells of the crystals,
  whose Brillouin zones were sampled by
  $8\times 8\times 6$ (for the crystals with wurtzite structure) or
  $8\times 8\times 8$ or $9\times 9\times 9$ (for the crystals with rocksalt and zincblende structures)
  {\bf k}-point meshes generated according to Monkhorst-Pack scheme,\cite{MP76}
  enabling us to achieve convergence of the energy within 1 meV/atom.
Using the optimized crystal structures,
  we then performed band-structure and density-of-states calculations
  in order to obtain the band gap $E_g$ and the $d$ band position $\varepsilon_d$, respectively.
Besides we performed geometry optimizations
  for the O$_2$ and S$_8$ molecules and the bulk solids of Se, Te, Zn, and Cd,
  and employed the respective equilibrium total energies
  in the computation of the formation energy $\Delta H_f$.

As indicated in Section~\ref{giris},
  we set the value of $U$ by reproducing the experimental value of the band gap in the HSE+$U$ calculations,
  which is justified in Section~\ref{sonuclar}.
Thus, we carried out the HSE+$U$ calculations for a range of $U$ values,
  and studied the calculated band gap as a function of $U$.
Since our results showed that
  the variation of the band gap with $U$ is virtually linear,
  we performed a linear fit to obtain the value of $U$
  that corresponds to the measured band gap.
The value of $U$ obtained via this procedure,
  which is {\it optimal} in reproducing the experimental value of the band gap,
  is denoted by $U^\ast$.
The HSE+$U$ calculation that yields the experimental, i.e., {\it targeted}, value of the band gap
  is named here as the HSE+$U^\ast$ calculation.

\begin{figure}[b]
  \begin{center}
    \resizebox{0.50\textwidth}{!}{%
      \includegraphics{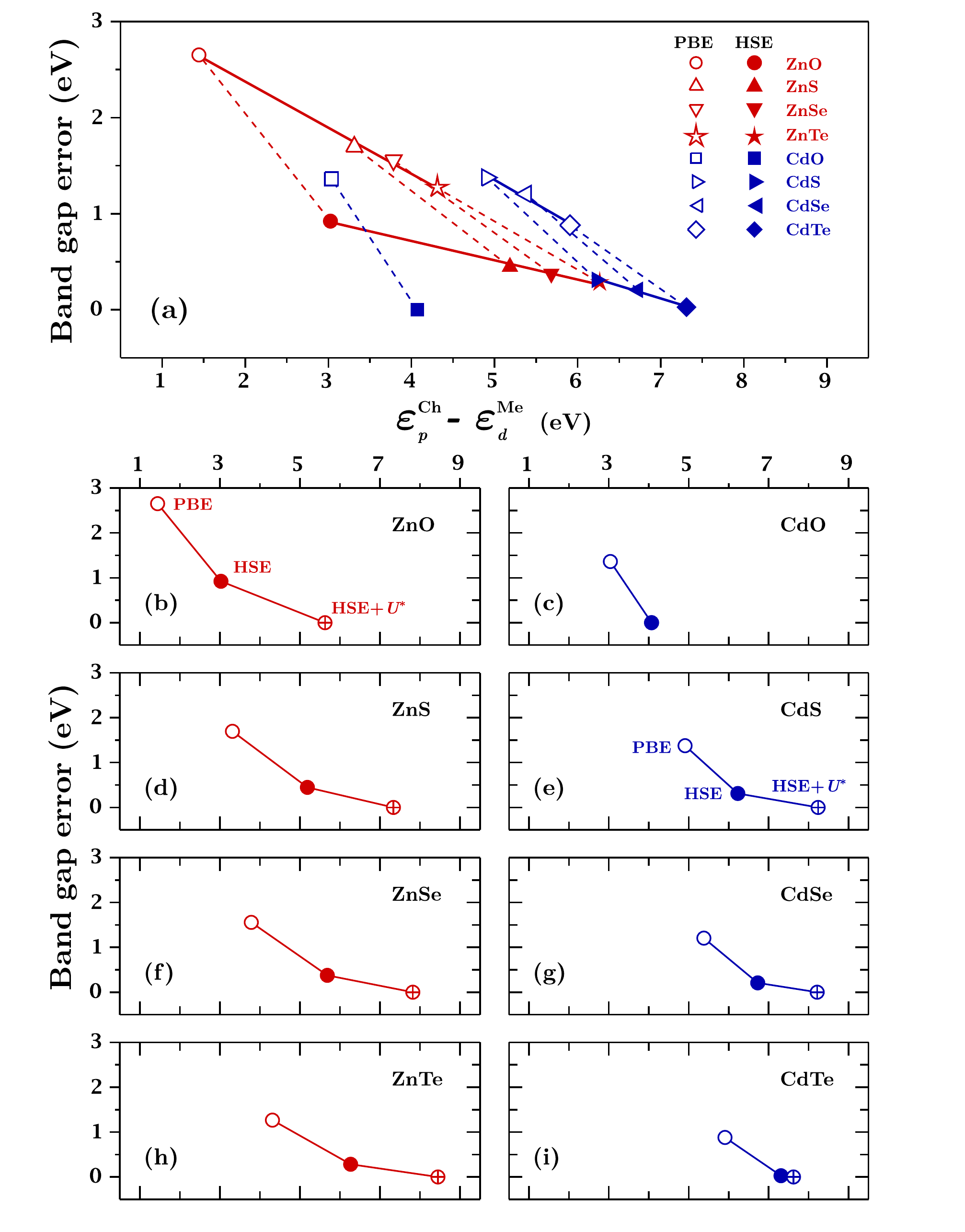}
    }
  \end{center}
  \vspace{-0.4cm}
  \caption{(Color online)
           The band gap error $\Delta E_g$ versus
           the difference $\Delta \varepsilon_{pd}=\varepsilon_{p}^{\rm Ch}-\varepsilon_{d}^{\rm Me}$
           for zinc and cadmium monochalcogenides.
           The PBE- and HSE-calculated values are marked by the empty and filled symbols, respectively, in the top-most panel (a).
           In the lower panels (b)-(i), the results of the combined HSE+$U^\ast$ ($\oplus$) calculations
             are presented together with those of the PBE (empty symbols) and HSE (filled symbols) calculations.
          }
   \label{dEgvsdepd}
\end{figure}

It should be mentioned that the HSE band energy differences
  depend on the value of the screening parameter $\omega$,
  which is not necessarily universal.
It was, however, demonstrated\cite{brothers08,janesko09} that $\omega=0.207$ \AA$^{-1}$
  as used in HSE06
  is an average optimal value
  for which the band energy differences
  approximate rather accurately {\it quasiparticle excitation energies},
  for a variety of semiconductors.
Therefore, the HSE band energy differences are often {\it directly} compared
  to the experimental band gaps\cite{henderson11}
  (e.g., in order to demonstrate\cite{heyd05} the success of the HSE calculations
  in reproducing the experimental band gaps).
In addition to this,
  as long as the HSE+$U$ approach could be regarded as an approximation to the $GW$ method,
  it would be preferential to use the quasiparticle energy differences (the $GW$-calculated band gaps)
  in our procedure for setting the value of $U^\ast$.
However, the $GW$-calculated band gaps are usually in good agreement with the experimental band gaps
  (e.g., Ref.~\onlinecite{shishkin07}).
It should, on the other hand, be also noted that the $GW$@HSE calculations {\it overestimate} the band gap
  of a number of semiconductors including CdS and ZnS (Ref.~\onlinecite{fuchs07}).
Hence, we preferred to utilize the experimental band gaps instead of the $GW$-calculated energy differences,
  which is also convenient from a practical point of view
  since it enables one to avoid performing quasiparticle calculations
  that might easily become computationally exhaustive, especially for large-scale (e.g., defect) calculations.

%
%
\section{\label{sonuclar}Results and Discussion}

We first quantify the relationship between the band gap error $\Delta E_g$ in the GGA and HSE calculations
  and the position of $d$ level
  in the case of zinc and cadmium chalcogenides
  since the latter is, in effect, adjusted by varying the value of $U$.
Figure~\ref{dEgvsdepd}(a) shows a plot of $\Delta E_g$
  versus 
  the difference $\Delta \varepsilon_{pd}=\varepsilon_{p}^{\rm Ch}-\varepsilon_{d}^{\rm Me}$,
  where $\varepsilon_{p}^{\rm Ch}$ and $\varepsilon_{d}^{\rm Me}$ denote the $p$- and $d$-state
  energies of the chalcogen and metal atoms, respectively.
In zinc and cadmium chalcogenides,
  the $d$ band is located below and next to the topmost valence band.\cite{supplemental}
Thus, the valence-band maximum turns out to be {\it above} its actual position
  if the metal $d$ states are positioned {\it too} high (as in both the GGA and HSE calculations),
  which contributes to the underestimation of the band gap.
The difference $\Delta \varepsilon_{pd}$ is therefore used here
  to quantify the relationship between the band gap error and the position of $d$ level.
In Fig.~\ref{dEgvsdepd}(a), a linear trend is noticeable for each set of data, cf. the solid lines,
  with the exception of data points for CdO.
It is seen that the band gap error is  proportional
  (with a negative slope) to $\Delta \varepsilon_{pd}$.
We obtain, via fitting,
\begin{eqnarray}
  \Delta E_g&=& -0.48~\Delta \varepsilon_{pd}+3.33 ~~~({\rm PBE})\nonumber \\
            &=& -0.20~\Delta \varepsilon_{pd}+1.52 ~~~({\rm HSE})
\end{eqnarray}
for Zn chalcogenides, and
\begin{eqnarray}
  \Delta E_g&=& -0.50~\Delta \varepsilon_{pd}+3.84 ~~~({\rm PBE})\nonumber \\
            &=& -0.26~\Delta \varepsilon_{pd}+1.97 ~~~({\rm HSE})
\end{eqnarray}
for Cd chalcogenides (excluding CdO),
where $\Delta E_g$ and $\Delta \varepsilon_{pd}$ are both in eV.
It is clear, comparing the data points represented by empty (PBE) and filled (HSE) symbols connected by dashed lines,
  that the band gap error is reduced
  when the difference between the chalcogen $p$- and metal $d$-state energies is increased.
This applies to all II-VI semiconductors studied here, including CdO.
As shown in Figs.~\ref{dEgvsdepd}(b)-\ref{dEgvsdepd}(i),
  $\Delta \varepsilon_{pd}$ is significantly increased in the HSE+$U^\ast$ calculations,
  making $\Delta E_g$ vanish.
This is reassuring that the optimal Hubbard parameter $U^\ast$ could be determined
  by matching the experimental band gap.

\begin{figure}
  \begin{center}
    \resizebox{0.50\textwidth}{!}{%
      \includegraphics{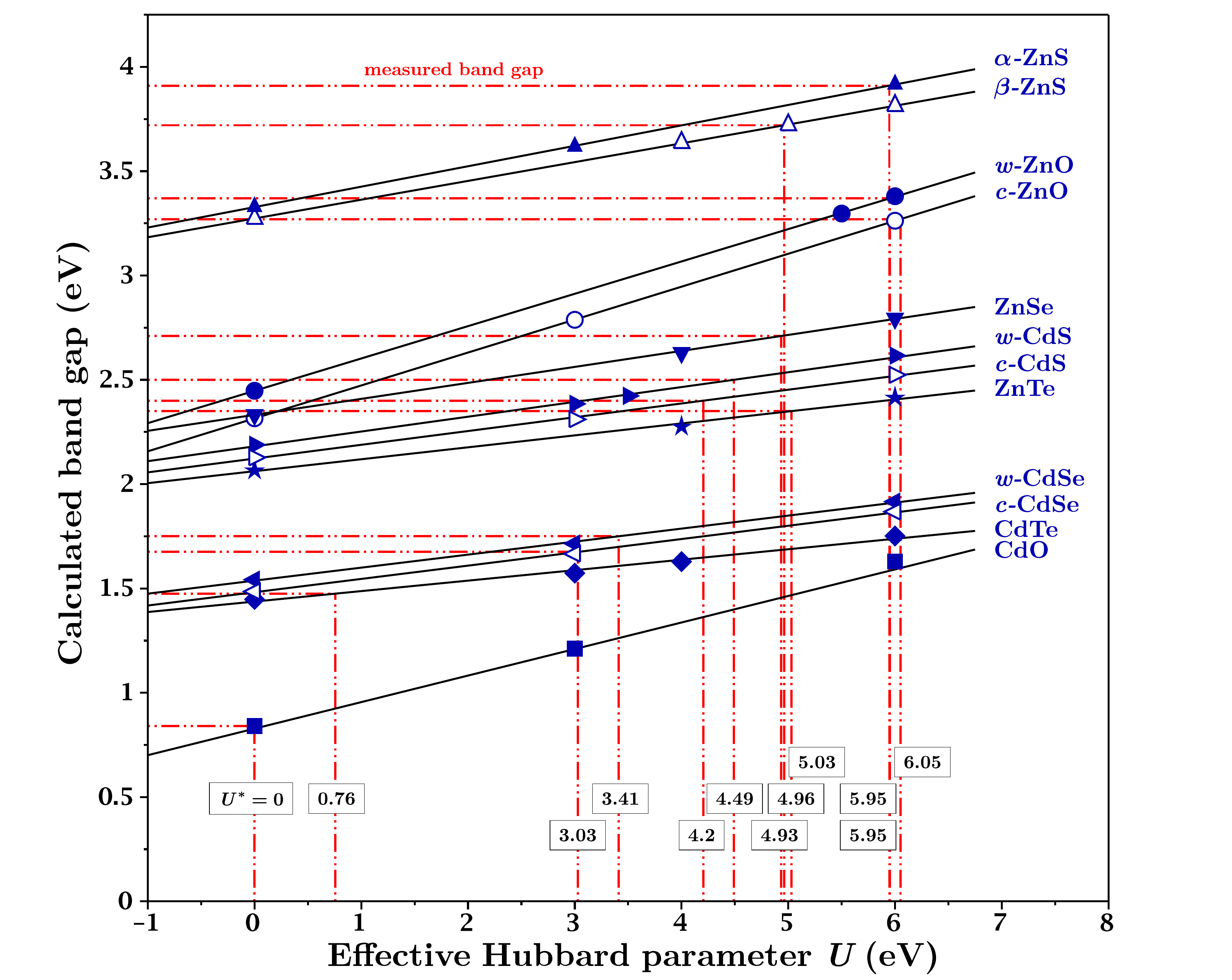}
    }
  \end{center}
  \vspace{-0.4cm}
  \caption{(Color online)
          The band gaps $E_g$ obtained in our HSE+$U$ calculations
          as a function of the effective Hubbard parameter $U$
          for zinc and cadmium monochalcogenides.
          The symbols represent the calculated $E_g$ values,
          and the solid lines connecting the symbols represent linear fits to the calculated points.
          The vertical dot-dashed lines
          mark the values for the optimal Hubbard parameter $U^\ast$ in eV,
          which correspond to the experimental $E_g$ values (marked by the horizontal dot-dashed lines).
          }
   \label{optU}
\end{figure}

\begin{table}
\caption{\label{tablo}
         The optimal Hubbard parameter $U^\ast$,
         the experimental band gap $E_g$, and
         the HSE band gap error $\Delta E_g^{\rm HSE}$ (all in eV)
         for zinc and cadmium monochalcogenides.
        }
\begin{ruledtabular}
\begin{tabular}{llccc}
Semiconductor & Crystal structure & $U$$^\ast$ & $E_g$  & $\Delta E_g^{\rm HSE}$ \\ \hline
CdO           & rocksalt          & 0.0        & 0.84   & 0.00                   \\
CdTe          & zincblende        & 0.8        & 1.48   & 0.03                   \\
$c$-CdSe      & zincblende        & 3.0        & 1.68   & 0.19                   \\
$w$-CdSe      & wurtzite          & 3.4        & 1.75   & 0.21                   \\
ZnTe          & zincblende        & 5.0        & 2.35   & 0.28                   \\
$c$-CdS       & zincblende        & 4.2        & 2.40   & 0.27                   \\
$w$-CdS       & wurtzite          & 4.5        & 2.50   & 0.31                   \\
ZnSe          & zincblende        & 5.0        & 2.71   & 0.38                   \\
$c$-ZnO       & zincblende        & 6.1        & 3.27   & 0.95                   \\
$w$-ZnO       & wurtzite          & 6.0        & 3.37   & 0.92                   \\
$\beta$-ZnS   & zincblende        & 5.0        & 3.72   & 0.45                   \\
$\alpha$-ZnS  & wurtzite          & 6.0        & 3.91   & 0.58                   
\end{tabular}
\end{ruledtabular}
\end{table}

\begin{figure}
  \begin{center}
    \resizebox{0.50\textwidth}{!}{%
      \includegraphics{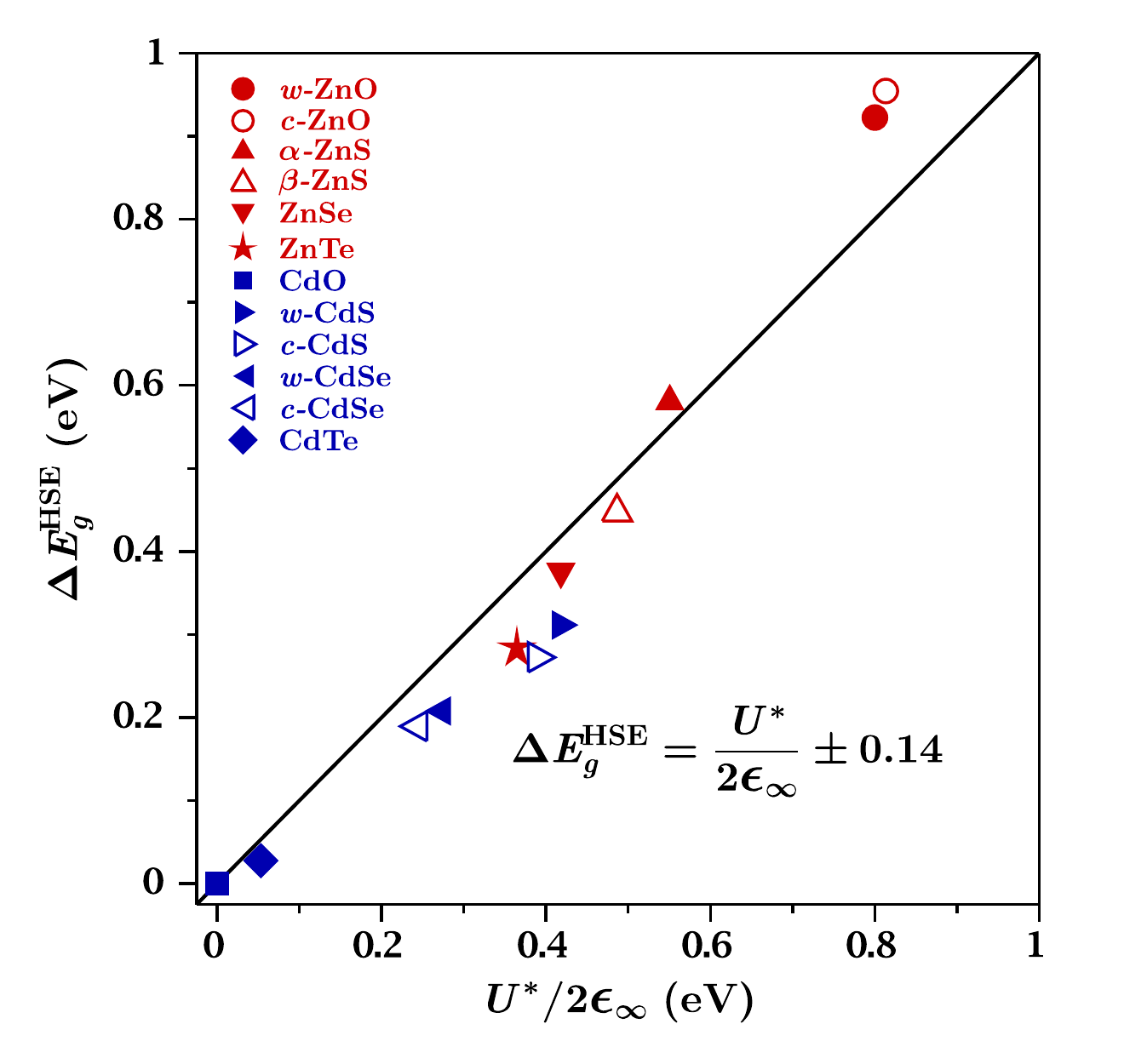}
    }
  \end{center}
  \vspace{-0.4cm}
  \caption{
          (Color online)
          The HSE band gap error $\Delta E_g^{\rm HSE}$ versus the ratio $U^\ast/2\epsilon_\infty$.
          }
   \label{UvsEg}
\end{figure}

\begin{figure*}
  \begin{center}
    \resizebox{1.00\textwidth}{!}{%
      \includegraphics{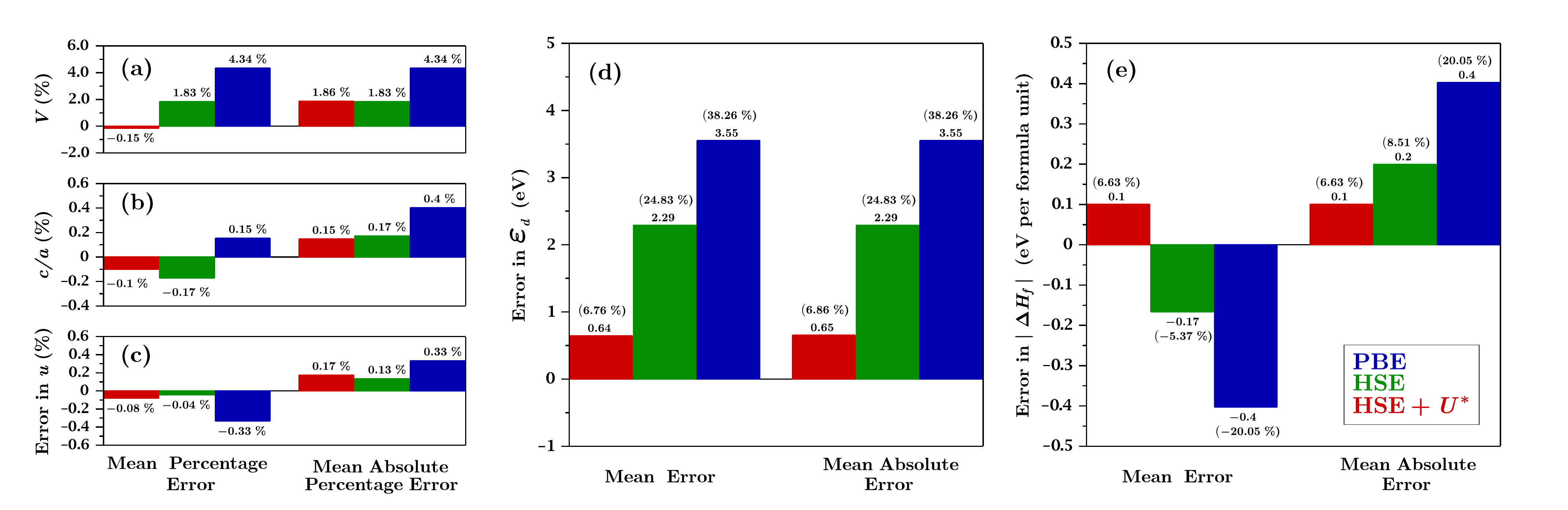}
    }
  \end{center}
  \vspace{-0.4cm}
  \caption{(Color online)
           A comparison of errors in the predictions via HSE+$U^\ast$ (red bars), 
                                                         HSE (green bars) and 
                                                         PBE (blue bars) calculations for
            the unit cell volume $V$ (a),
            the ratio $c/a$ of lattice parameters $a$ and $c$ (b),
            the internal lattice parameter $u$ (c),
            the $d$ band position $\varepsilon_d$ (d), and
            the absolute value of formation energy $\left | \Delta H_f \right |$ (e).
          }
   \label{karsilastir}
\end{figure*}

We now determine the $U^\ast$ values that corresponds to vanishing $\Delta E_g$
  for the II-VI semiconductors under consideration.
Thus, the results of HSE+$U$ calculations for a range of $U$ values
  are given in Fig.~\ref{optU} where the calculated band gap is plotted as a function of $U$.
Note that the variation of the band gap with the effective Hubbard parameter is virtually linear
  (with a different slope for each system).
For each compound, a linear fit is thus performed,
  which yields the solid lines in Fig.~\ref{optU}.
The $U^\ast$ values are marked by vertical dot-dashed lines,
  which correspond to the measured band gap (marked by horizontal dot-dashed lines).
Table~\ref{tablo} gives
  the optimal Hubbard parameters and corresponding band gaps
  for zinc and cadmium monochalcogenides.
It should be remarked
  that one obtains $U^\ast = 0$ for CdO
  since the measured value of the band gap of CdO is reproduced already in the HSE calculation, as mentioned in Section~\ref{giris}.

Next we compare the values of $\Delta E_g$ and $\Delta \varepsilon_{pd}$
  obtained in the HSE+$U^\ast$ calculations to those obtained in the PBE and HSE calculations.
Figures~\ref{dEgvsdepd}(b)-\ref{dEgvsdepd}(i)
  show a plot of the band gap error $\Delta E_g$ versus the difference $\Delta \varepsilon_{pd}$
  for the II-VI semiconductors under consideration.
As already noted, the HSE calculations yield
  an increased value for $\Delta \varepsilon_{pd}$
  in association with a reduced band gap error,
  in comparison to the PBE calculations.
The difference $\Delta \varepsilon_{pd}$ is further increased in the HSE+$U$ calculations,
  reducing the band gap error further.
Having $U=U^\ast$ in this trend makes $\Delta E_g$ vanish, with adequate increase of $\Delta \varepsilon_{pd}$.

It is seen in Table~\ref{tablo}
  that the larger $E_g$ the greater $U^\ast$ (with few exceptions).
This implies that employing a large (small) $U^\ast$ would be necessary
  for a wide (narrow) band gap semiconductor for which 
  the HSE band gap error $\Delta E_g^{\rm HSE}$ is rather large (small), cf. Figure~\ref{Egcalvsexp}(a).
Thus, having a large band gap error in the HSE calculation
  necessitates using a large $U^\ast$ for correction.
Furthermore,
  there appears to be a roughly monotonic relationship
  between $U^\ast$ and $\Delta E_g^{\rm HSE}$, cf. Table~\ref{tablo}.
Our analysis presented in Fig.~\ref{UvsEg}
  shows that
  this relationship could be quantified
  by taking into account the screening effects through the high-frequency dielectric constant $\epsilon_\infty$.
A plot of $\Delta E_g^{\rm HSE}$ versus $U^\ast/2\epsilon_\infty$ is given in
  Fig.~\ref{UvsEg}
  where all data points satisfy
  \begin{equation}
    \Delta E_g^{\rm HSE}= \frac{U^\ast}{2\epsilon_\infty} \pm 0.14 ~{\rm eV}.
    \label{UvsdEg}
  \end{equation}
Here both $U^\ast$ and $\Delta E_g^{\rm HSE}$ are in eV.
Note that the shift in the occupied (unoccupied) $d$ state energies
  due to the $U^\ast$ term would be
  $-U^\ast/2$ ($U^\ast/2$) if the hybridization and screening effects are ignored.\cite{anisimov93}
Thus, the correction to the band gap would be proportional to $U^\ast/2$,
  ignoring the dielectric screening,
  for the II-VI semiconductors studied here
  since their lower conduction bands have virtually no contribution from the metal $d$ states.\cite{supplemental}
On the other hand, the band gap correction needs to be scaled by $\epsilon_\infty$
  in order to reflect the dielectric screening of the Coulomb potential in a solid.\cite{harrison85}
Thus, the $U^\ast$ term added to the hybrid (HSE) functional
  results in a correction of $U^\ast/2\epsilon_\infty$ to the band gap.
This explanation justifies our means of setting the value of $U^\ast$
  by matching the experimental band gap.
It also implies that an approximate value for the optimal Hubbard parameter
  could {\it a priori} be obtained by inverting Eq.~(\ref{UvsdEg}), i.e.,
  $U^\ast \approx 2\epsilon_\infty(E_g - E_g^{\rm HSE})$, provided that
  the experimental and HSE-calculated band gaps $E_g$ and $E_g^{\rm HSE}$
  as well as the high-frequency dielectric constant $\epsilon_\infty$
  are available.
Note that the hybrid-functional calculations could be utilized
  to obtain $\epsilon_\infty$
  when the experimental data is not available, cf. Table~I of Ref.~\onlinecite{paier08}.

It is interesting to point out that one could assign a single $U^\ast$ value of $\sim 5$~eV for ZnTe, ZnSe and $\beta$-ZnS
  while $U^\ast \sim 6$~eV for $c$-ZnO, $w$-ZnO and $\alpha$-ZnS, cf. Table~\ref{tablo}.
Thus, a mean value of $U^\ast_{\rm Zn} \approx 5.5$~eV appears to be adequate
  for {\it all} Zn compounds studied here.
It is clearly pleasing to obtain a single (universal) $U^\ast$ value for Zn,
  which is almost independent of the composition or crystal structure of the relevant zinc compounds,
  for its practical importance
  since it would allow one to set $U^\ast_{\rm Zn} \approx 5.5$~eV
  in the studies on {\it alloyed} systems made of Zn, O, S, Se, Te atoms.

In order to assess the improvement of the HSE+$U$ approach
  in relation to the general physical description of the foregoing semiconductors,
  we computed the mean error in 
  (i) the optimized crystal structures,
  (ii) the $d$ band positions, and 
  (iii) the formation energies
  of the metal chalcogenides under consideration.
Accordingly, a comparison of errors in the predictions of the HSE+$U^\ast$, HSE and PBE calculations is presented
  Fig.~\ref{karsilastir} where the comparison is performed
  for the unit cell volume $V$                                     [in Fig.~\ref{karsilastir}(a)],
  for the ratio $c/a$ of (wurtzite) lattice parameters $a$ and $c$ [in Fig.~\ref{karsilastir}(b)],
  for the internal parameter $u$ of wurtzite structure             [in Fig.~\ref{karsilastir}(c)],
  for the $d$ band position $\varepsilon_d$                        [in Fig.~\ref{karsilastir}(d)], and
  for the formation energy $\Delta H_f$                            [in Fig.~\ref{karsilastir}(e)].
Our analysis reveals the following:

First, we see in Figs.~\ref{karsilastir}(a)-(c) that
  the optimization of the crystal structure
  via HSE or HSE+$U^\ast$ calculation results in a similarly more accurate description,
  in comparison to the PBE calculations.
Thus, the HSE+$U^\ast$ calculations seem to preserve the accuracy of the HSE calculations
  in the crystal structure optimizations.

Secondly, Fig.~\ref{karsilastir}(d) shows that there is a significant correction
  to the $d$ band position thanks to adding $U^\ast$ term to the HSE functional:
The mean error in the $\varepsilon_d$ prediction becomes $\sim$~0.6 eV
  in the HSE+$U^\ast$ calculations,
  compared to $\sim$~2.3 (3.6) eV in the HSE (PBE) calculations.
It should also be noted
  that the variation of the difference  $\Delta \varepsilon_d^\ast = \varepsilon_d^{{\rm HSE}+U^\ast}- \varepsilon_d^{\rm HSE}$
  with $U^\ast$ is almost linear,\cite{ikinci}
  which is consistent with $\Delta \varepsilon_d^\ast \approx -0.35~U^\ast $,
  where both $\Delta \varepsilon_d^\ast$ and $U^\ast $ are in eV.
Thus, using a larger $U^\ast$ yields a larger correction to $\varepsilon_d$,
  shifting the $d$ band to a lower position that is closer to its experimental location.
Recall that employing a larger $U^\ast$ is necessary for the systems
  with a larger HSE band gap error (cf. Table~\ref{tablo}).
Hence, the improvement in predicting the $d$ band position
  via HSE+$U^\ast$ calculations
  is warranted
  {\it since} the value of $U^\ast$ is determined by matching the experimental band gap.

Finally, as for the improvement of the HSE+$U$ approach in the prediction of formation energies,
  Fig.~\ref{karsilastir}(e) shows that
  the mean absolute error in $\left | \Delta H_f \right |$
  is on the order of $\sim$~0.1, 0.2, and 0.4~eV per formula unit
  in the HSE+$U^\ast$, HSE, and PBE calculations, respectively.
Thus, the HSE+$U^\ast$ calculations
  result in a more accurate description of crystal energetics of zinc and cadmium monochalcogenides,
  compared to the HSE and PBE calculations.
Note that the mean error in $\left | \Delta H_f \right |$
  turns out to be {\it positive} in the HSE+$U^\ast$ calculations,
  which is {\it negative} in the HSE calculations.
This indicates that the error in the formation energies
  could be further reduced,
  whenever necessary,
  by re-adjusting the value of $U^\ast$.

%
%
\section{\label{netice}Conclusion}

In this work,
  we treated the hybrid functional scheme and the DFT+$U$ method as complementary
  rather than alternative approaches
  in studying a set of II-VI semiconductors with localized $d$ states.
This led us to introduce the HSE+$U$ approach
  where the range-separated HSE hybrid functional is combined with the Hubbard $U$.
Furthermore, we regarded $U$ as a semiempirical parameter.
This enabled us to determine an optimal value $U^\ast$ of the
  Hubbard parameter, for which the HSE+$U$ calculation
  yields a {\it targeted} (e.g., experimental) value of the band gap.
We find that the correction to the band gap
  due to the additional $U^\ast$ term is roughly given by
  $U^\ast/2\epsilon_\infty$,
  which is in line with theoretical reasoning.
The results of a variety of HSE+$U^\ast$ calculations
  performed for zinc and cadmium monochalcogenides, viz., a subset of the semiconductors with localized $d$ states,
  indicate that
  an improved description of the electronic structure as well as crystal structure and energetics
  is obtained
  in these calculations,
  compared to the hybrid functional calculations employing the HSE functional without an additional Hubbard term.
The present study thus shows that
  adding the $U^\ast$  term to the HSE functional leads to more accurate prediction of both the electronic and crystal structures
  of II-VI semiconductors with localized states.

%
%
\begin{acknowledgments}
  The numerical calculations reported here were carried out at the High Performance and Grid Computing Center (TRUBA Resources) of TUBITAK ULAKBIM.
\end{acknowledgments}

%
%
\appendix*

\section{Band gap error in DFT and hybrid-functional calculations}

\begin{figure}
  \begin{center}
    \resizebox{0.50\textwidth}{!}{%
      \includegraphics{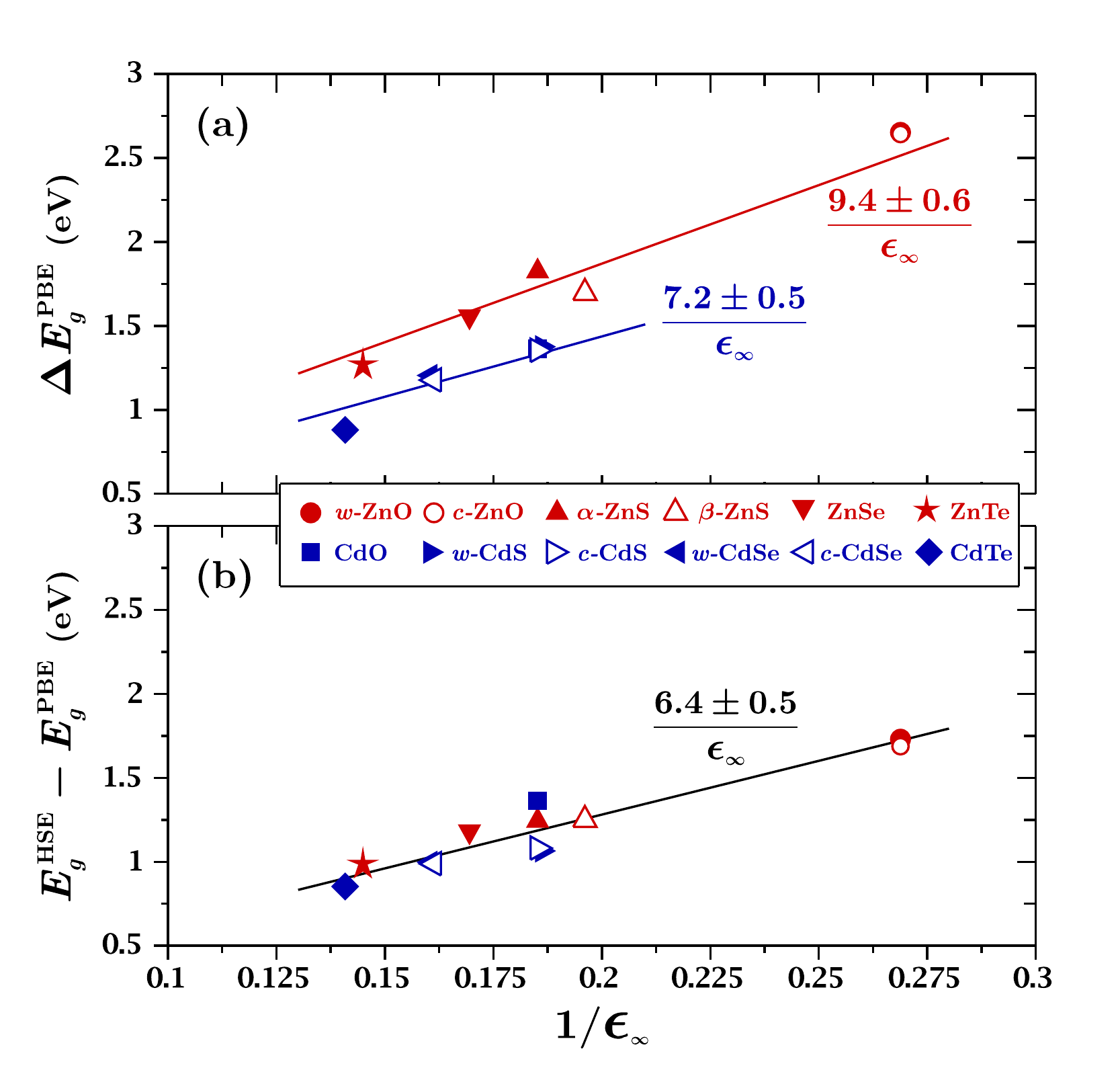}
    }
  \end{center}
  \vspace{-0.4cm}
  \caption{
          (Color online)
           The band gap error $\Delta E_g^{\rm PBE}$ in the PBE calculations (a) and
           the difference $E_g^{\rm HSE}-E_g^{\rm PBE}$ HSE- and PBE-calculated band gaps (b)
           versus the inverse dielectric constant $1/\epsilon_\infty$.
           The experimental values of $E_g$ and $\epsilon_\infty$
           are taken from Refs.~\onlinecite{bandgap1,bandgap2,bandgap3} and
                          Refs.~\onlinecite{adachi04,finkenrath67}, respectively.
          }
   \label{dEg}
\end{figure}

Figure~\ref{dEg}(a) shows a plot of the band gap error $\Delta E_g^{\rm PBE}=E_g-E_g^{\rm PBE}$ in the GGA calculation
  versus
  the inverse high-frequency dielectric constant $1/\epsilon_\infty$,
  where a nearly linear trend is noticeable for each set of data.
It is clear that 
  the band gap error is larger for materials with smaller $\epsilon_\infty$.
This is in line with the finding\cite{fiorentini95,linden86} that
  the self-energy correction to the DFT-calculated band gap
  is inversely proportional to the high-frequency dielectric constant.
Employing the LDA-calculated band gaps,
  it was found\cite{fiorentini95} that
  the product $\epsilon_\infty \Delta E_g^{\rm LDA} \approx 9.1\pm0.2$~eV is a material-independent constant.
We find that the product $\epsilon_\infty  \Delta E_g^{\rm PBE} = A$ is also roughly a constant,
  but with a different value for each class of systems:
  $A_{\rm Zn}=9.4 \pm 0.6$~eV and $A_{\rm Cd}=7.2 \pm 0.5$~eV for Zn and Cd chalcogenides, respectively.
Note that the data points in red (blue) in Fig.~\ref{dEg}(a) are consistent
  with the red (blue) line given by $\Delta E_g^{\rm PBE}=A_{\rm Zn}/\epsilon_\infty$ ($\Delta E_g^{\rm PBE}=A_{\rm Cd}/\epsilon_\infty$).
On the other hand,
  our results presented in Fig.~\ref{dEg}(b)
  show that the difference $E_g^{\rm HSE}-E_g^{\rm PBE}$
  is also inversely proportional to $\epsilon_\infty$
  so that $E_g^{\rm HSE}-E_g^{\rm PBE} \approx A^\prime /\epsilon_\infty $,
          where $A^\prime = 6.4 \pm 0.5$~eV is a material-independent constant.
Combining $\Delta E_g^{\rm PBE}-\Delta E_g^{\rm HSE}=E_g^{\rm HSE}-E_g^{\rm PBE} \approx A^\prime /\epsilon_\infty$ [cf. Fig.~\ref{dEg}(b)]
  and $\Delta E_g^{\rm PBE} \approx A/\epsilon_\infty$ [cf. Fig.~\ref{dEg}(a)],
  we obtain $\Delta E_g^{\rm HSE} \approx (A - A^\prime) /\epsilon_\infty$.
Note that both $A_{\rm Zn}$ and $A_{\rm Cd}$ are greater than $A^\prime$.
Thus,
  the band gap is underestimated in the HSE calculations in proportionality with $1/\epsilon_\infty$
  albeit there is a significant improvement in comparison to the respective GGA calculations.
%
%
%


\section*{Supplemental Material}

\setcounter{page}{1}
\renewcommand{\thepage}{S-\arabic{page}}
\renewcommand{\thefigure}{S\arabic{figure}}

The total (DOS) and projected (PDOS) density of states of zinc and cadmium monochalcogenides,
obtained via density-functional (PBE), hybrid-functional (HSE), and combined HSE+$U^\ast$ calculations,
are given Figs.~S1-S12.
Figure~S13 shows a plot of the difference $\Delta \varepsilon_d^\ast = \varepsilon_d^{{\rm HSE}+U^\ast}- \varepsilon_d^{\rm HSE}$
          versus the optimal Hubbard parameter $U^\ast$.

\begin{figure*}
  \begin{center}
    \resizebox{0.94\textwidth}{!}{%
      \includegraphics{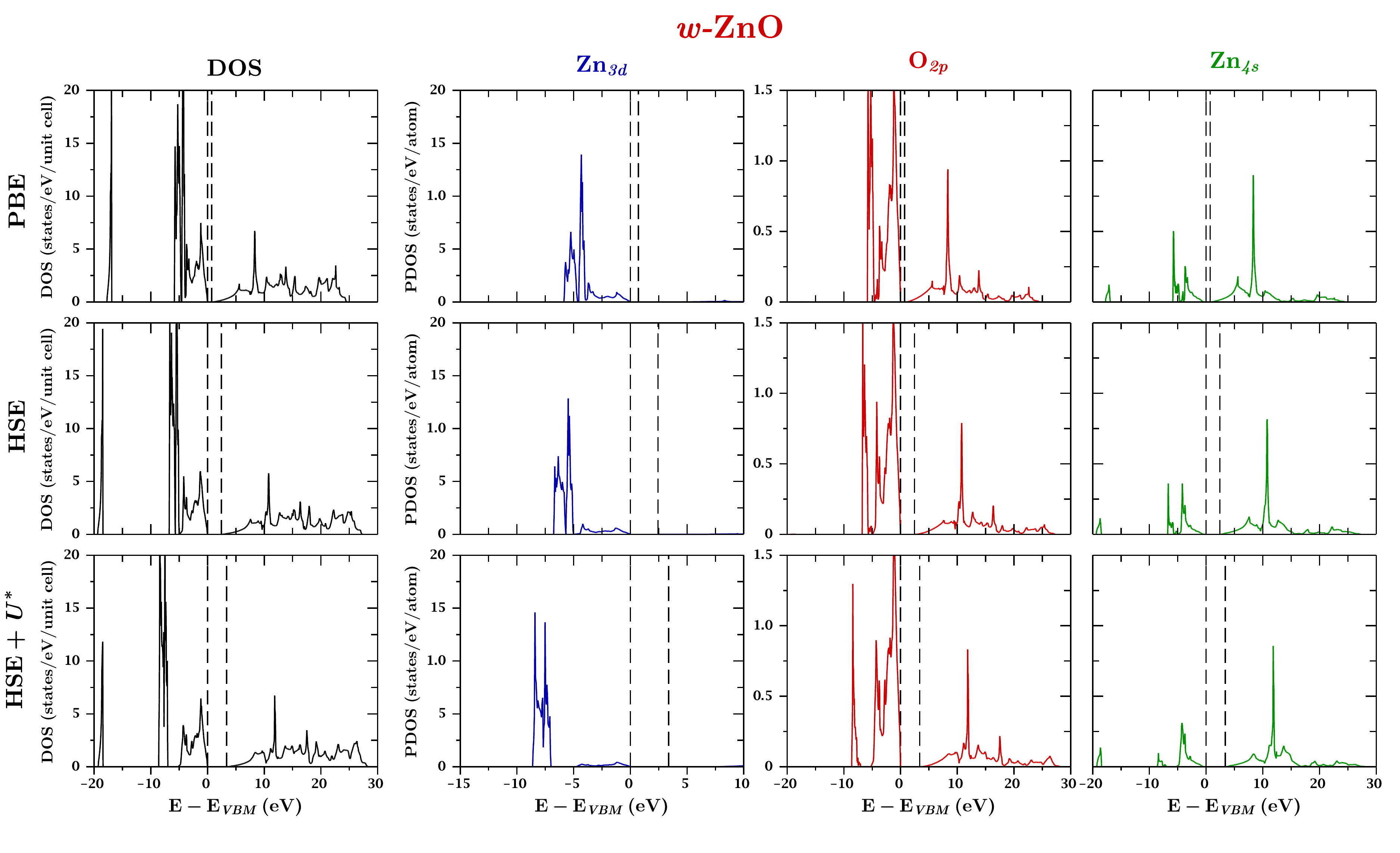}
    }
  \end{center}
  \caption{
          The total and projected density of states of $w$-ZnO.
          }
\end{figure*}

\begin{figure*}
  \begin{center}
    \resizebox{0.94\textwidth}{!}{%
      \includegraphics{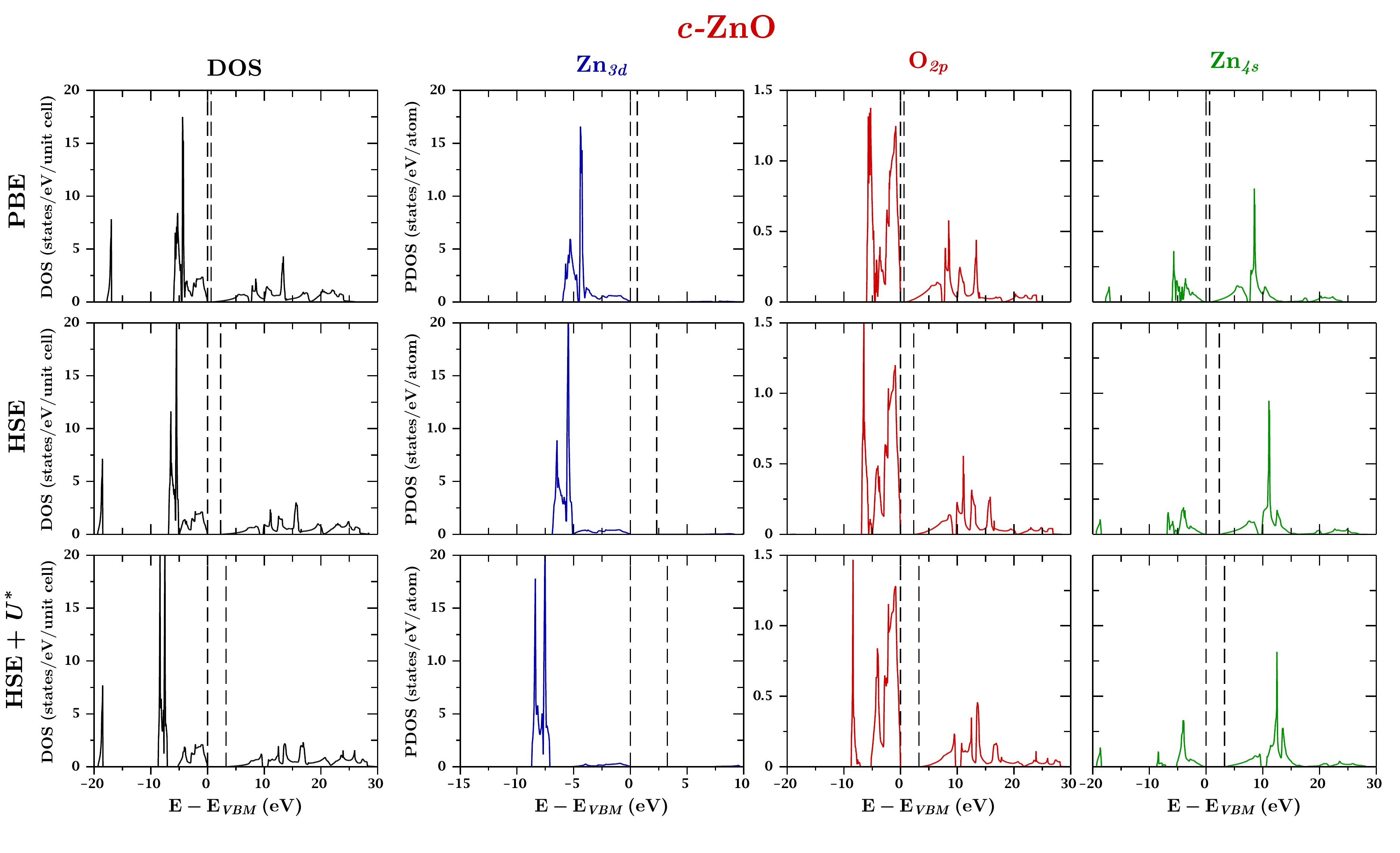}
    }
  \end{center}
  \caption{
          The total and projected density of states of $c$-ZnO.
          }
\end{figure*}

\begin{figure*}[landscape]
  \begin{center}
    \resizebox{0.94\textwidth}{!}{%
      \includegraphics{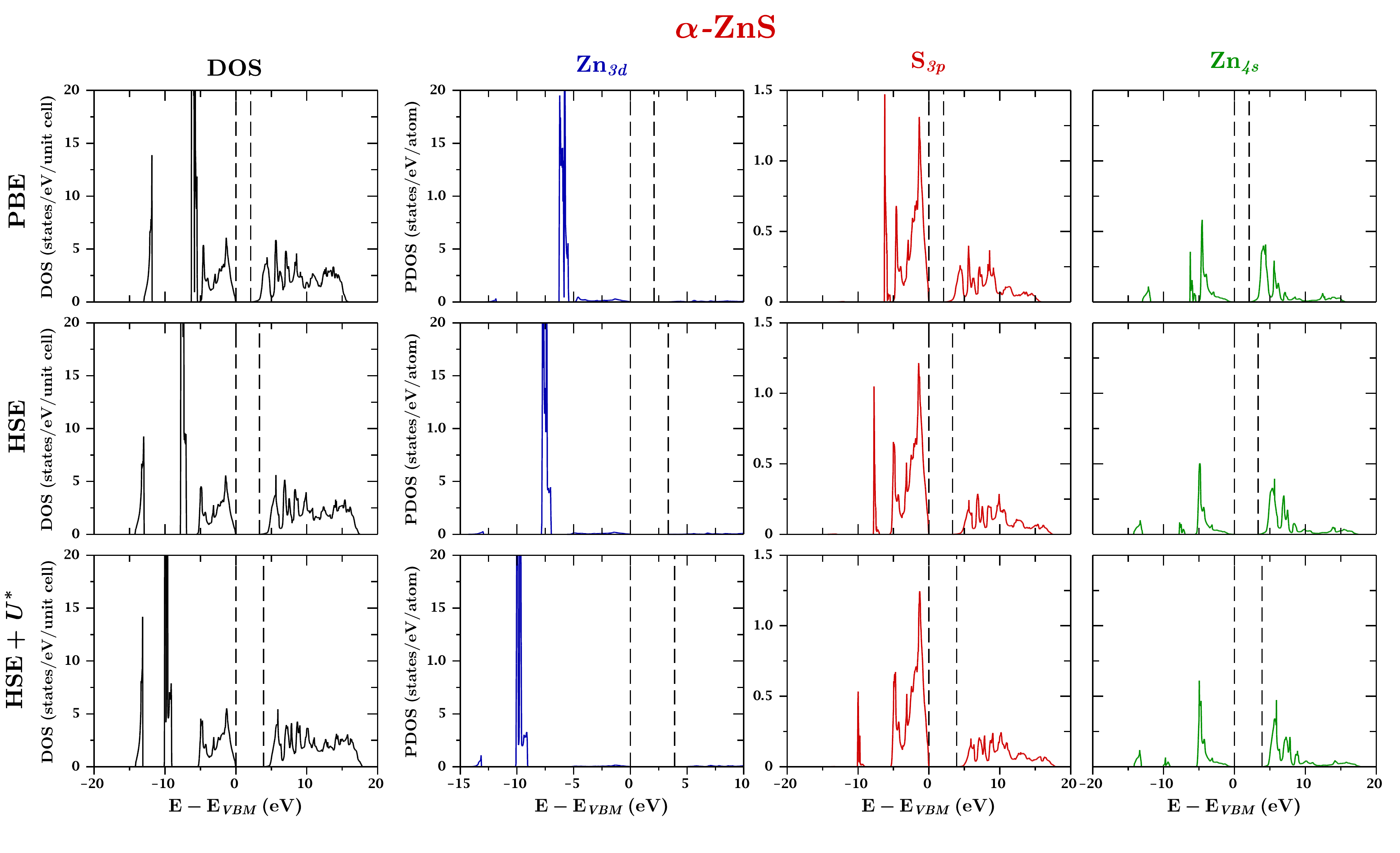}
    }
  \end{center}
  \caption{
          The total and projected density of states of $\alpha$-ZnS.
          }
\end{figure*}

\begin{figure*}
  \begin{center}
    \resizebox{0.94\textwidth}{!}{%
      \includegraphics{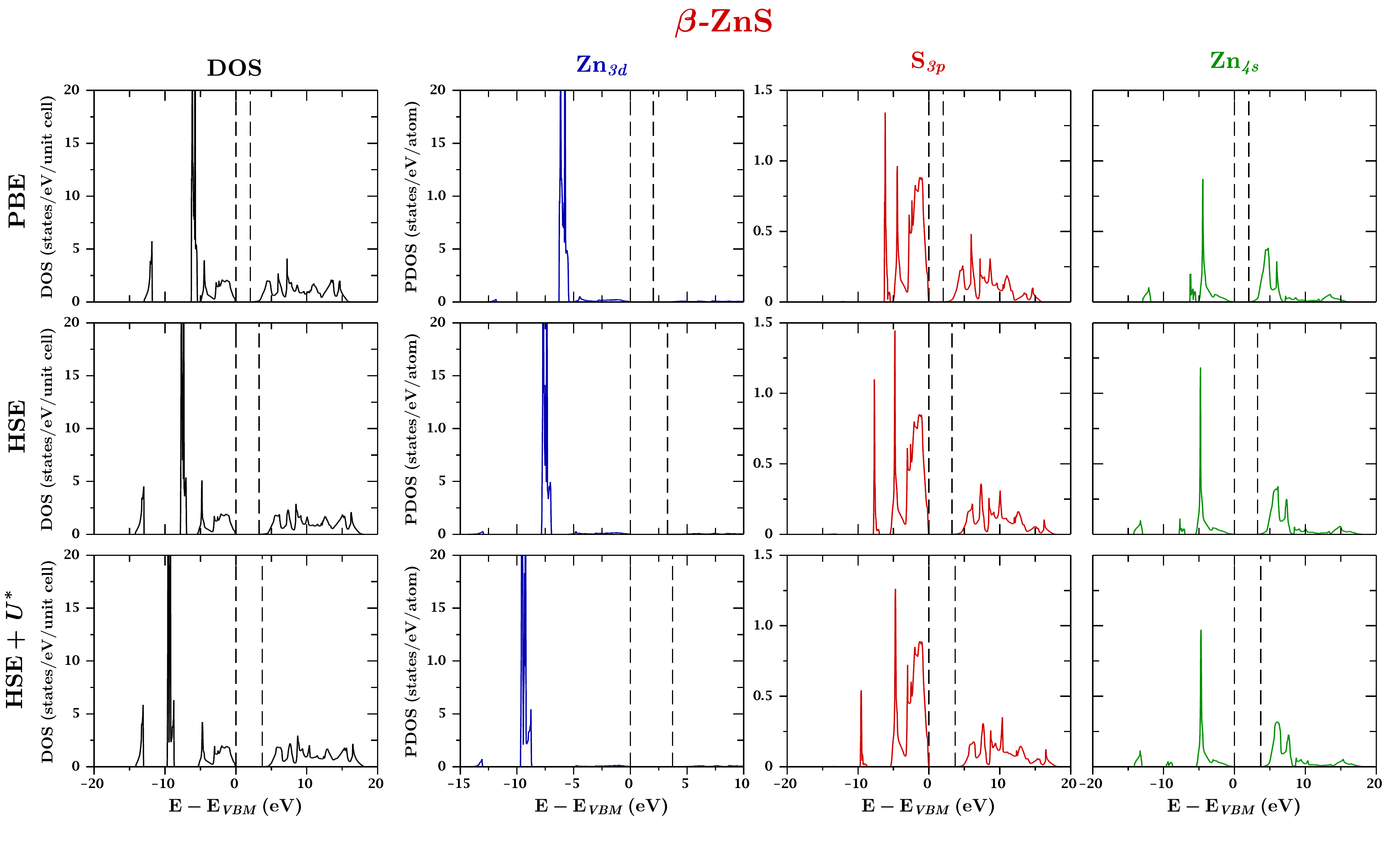}
    }
  \end{center}
  \caption{
          The total and projected density of states of $\beta$-ZnS.
          }
\end{figure*}

\begin{figure*}
  \begin{center}
    \resizebox{0.94\textwidth}{!}{%
      \includegraphics{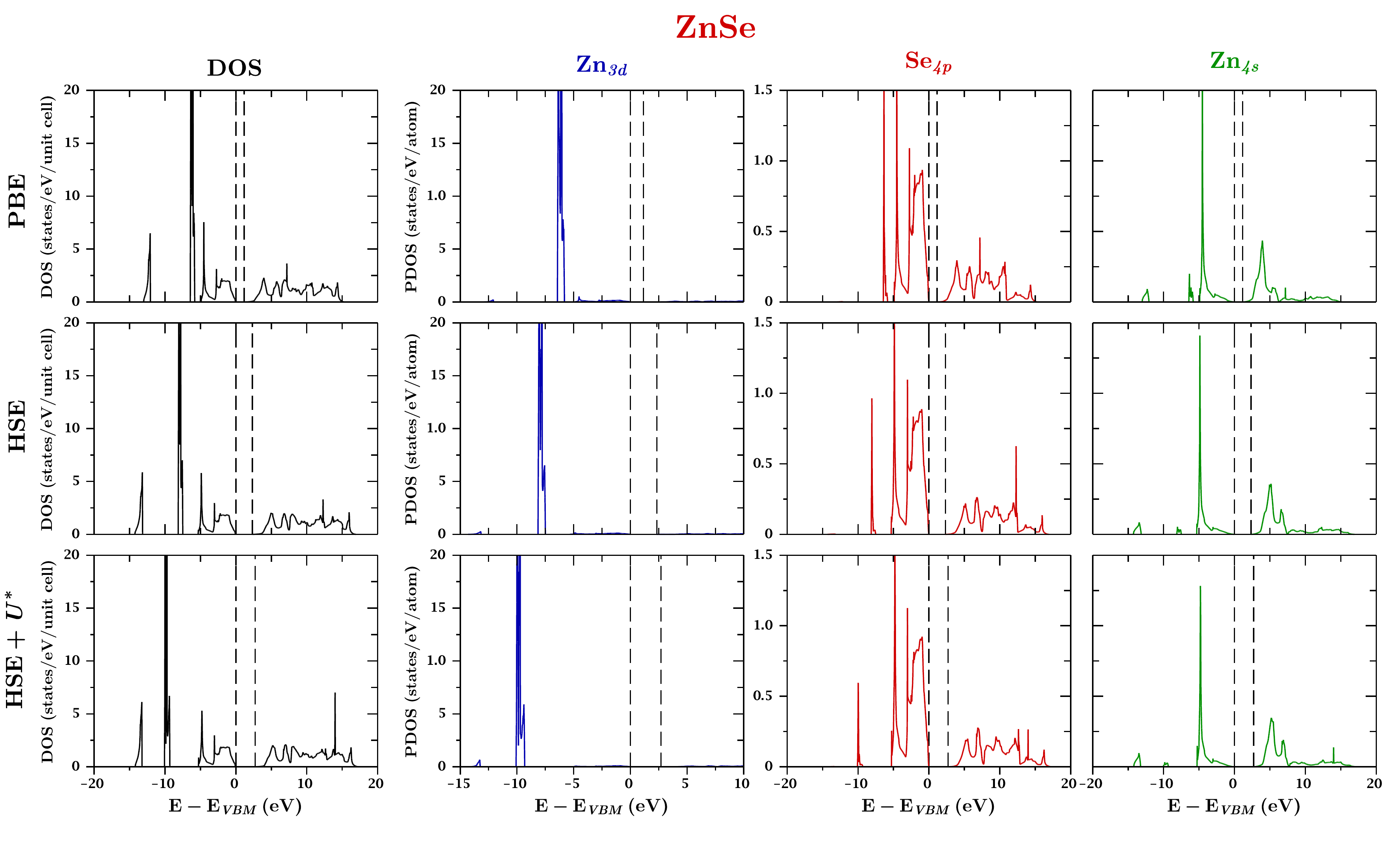}
    }
  \end{center}
  \caption{
          The total and projected density of states of ZnSe.
          }
\end{figure*}

\begin{figure*}
  \begin{center}
    \resizebox{0.94\textwidth}{!}{%
      \includegraphics{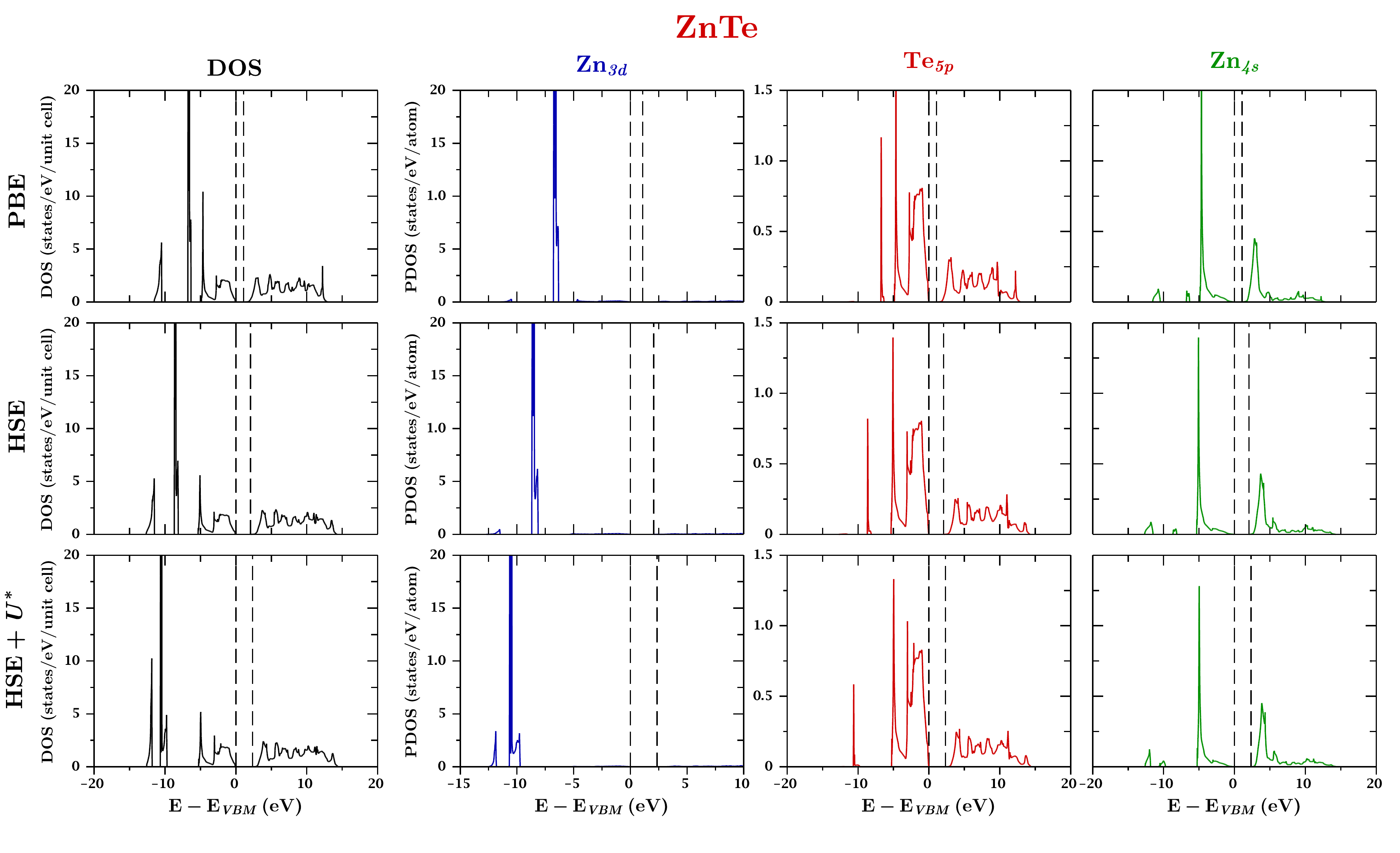}
    }
  \end{center}
  \caption{
          The total and projected density of states of ZnTe.
          }
\end{figure*}

\begin{figure*}
  \begin{center}
    \resizebox{0.94\textwidth}{!}{%
      \includegraphics{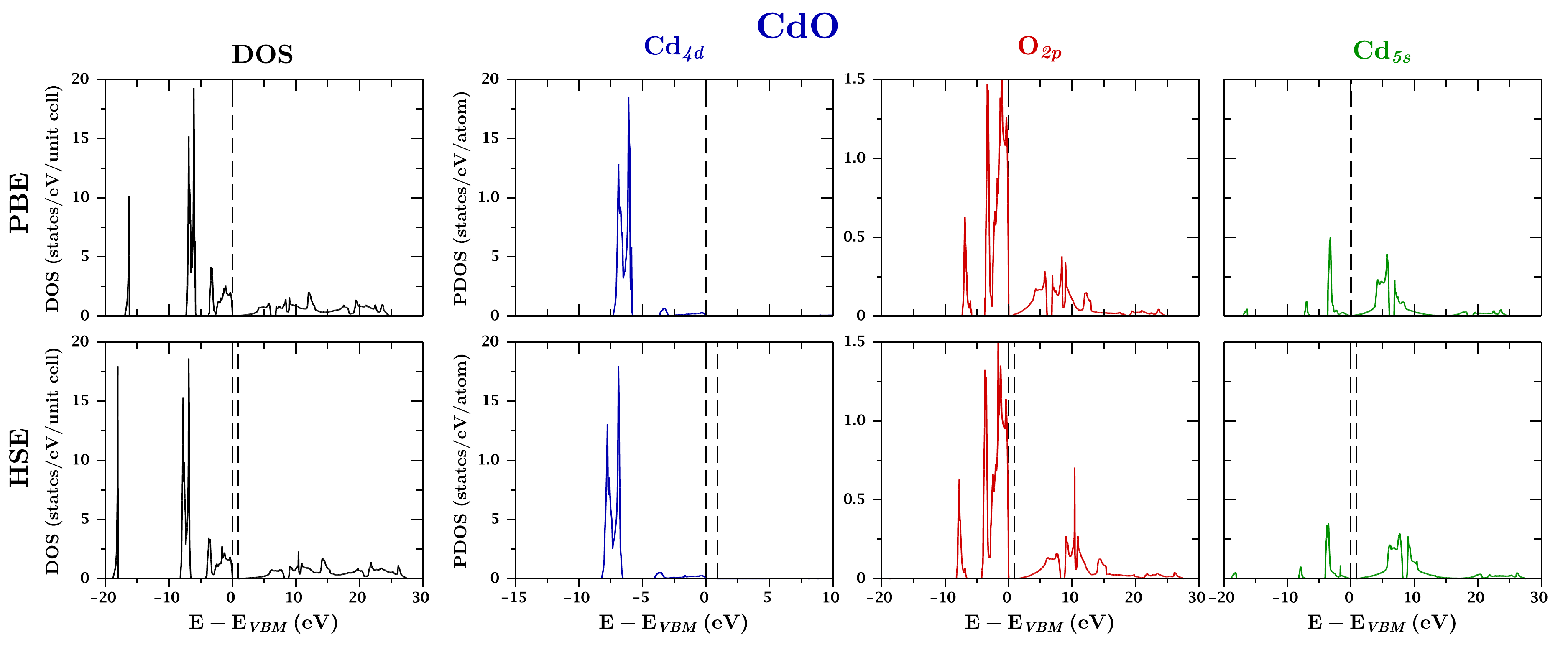}
    }
  \end{center}
  \caption{
          The total and projected density of states of CdO.
          }
\end{figure*}

\begin{figure*}
  \begin{center}
    \resizebox{0.94\textwidth}{!}{%
      \includegraphics{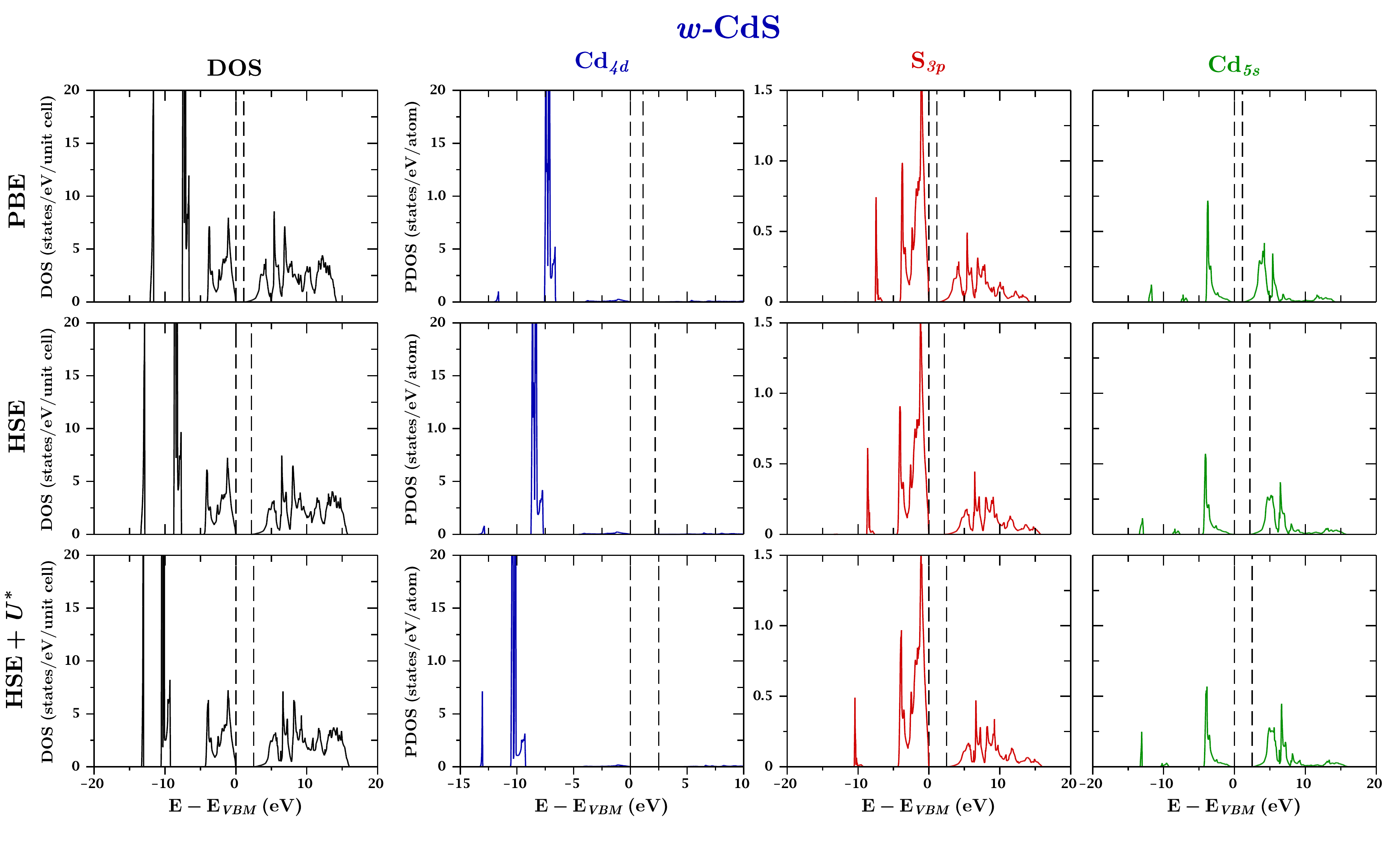}
    }
  \end{center}
  \caption{
          The total and projected density of states of $w$-CdS.
          }
\end{figure*}

\begin{figure*}
  \begin{center}
    \resizebox{0.94\textwidth}{!}{%
      \includegraphics{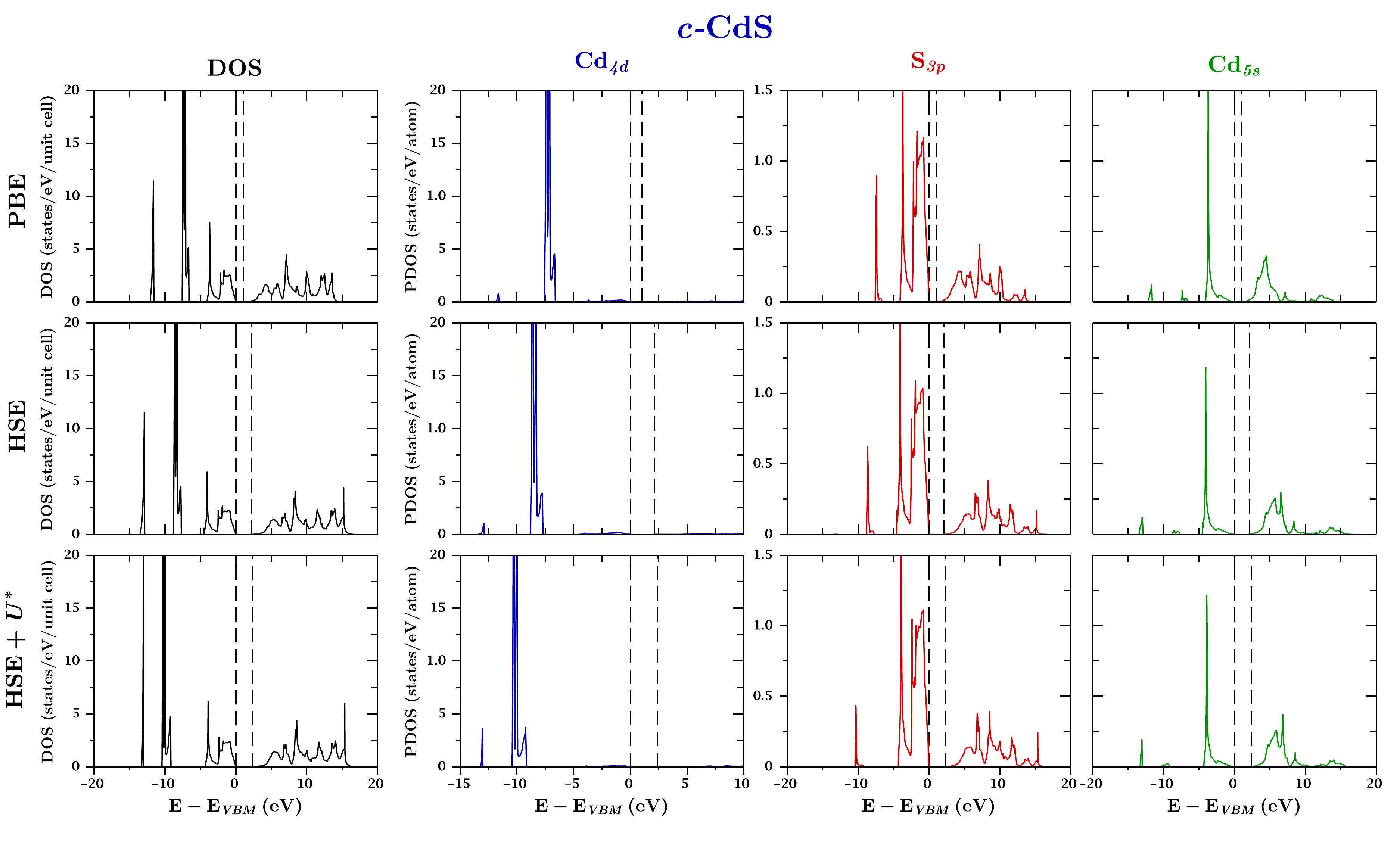}
    }
  \end{center}
  \caption{
          The total and projected density of states of $c$-CdS.
          }
\end{figure*}

\begin{figure*}
  \begin{center}
    \resizebox{0.94\textwidth}{!}{%
      \includegraphics{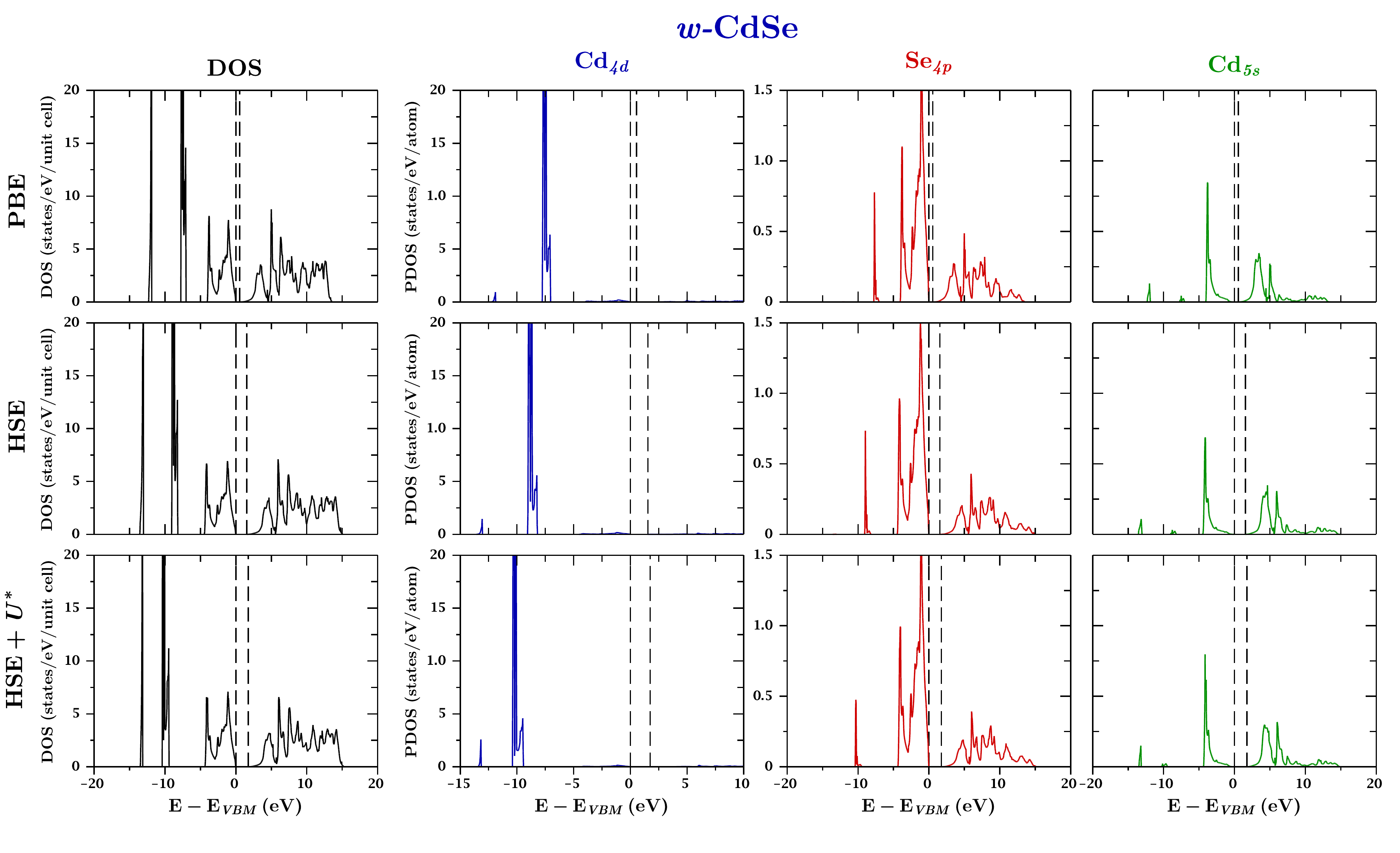}
    }
  \end{center}
  \caption{
          The total and projected density of states of $w$-CdSe.
          }
\end{figure*}

\begin{figure*}
  \begin{center}
    \resizebox{0.94\textwidth}{!}{%
      \includegraphics{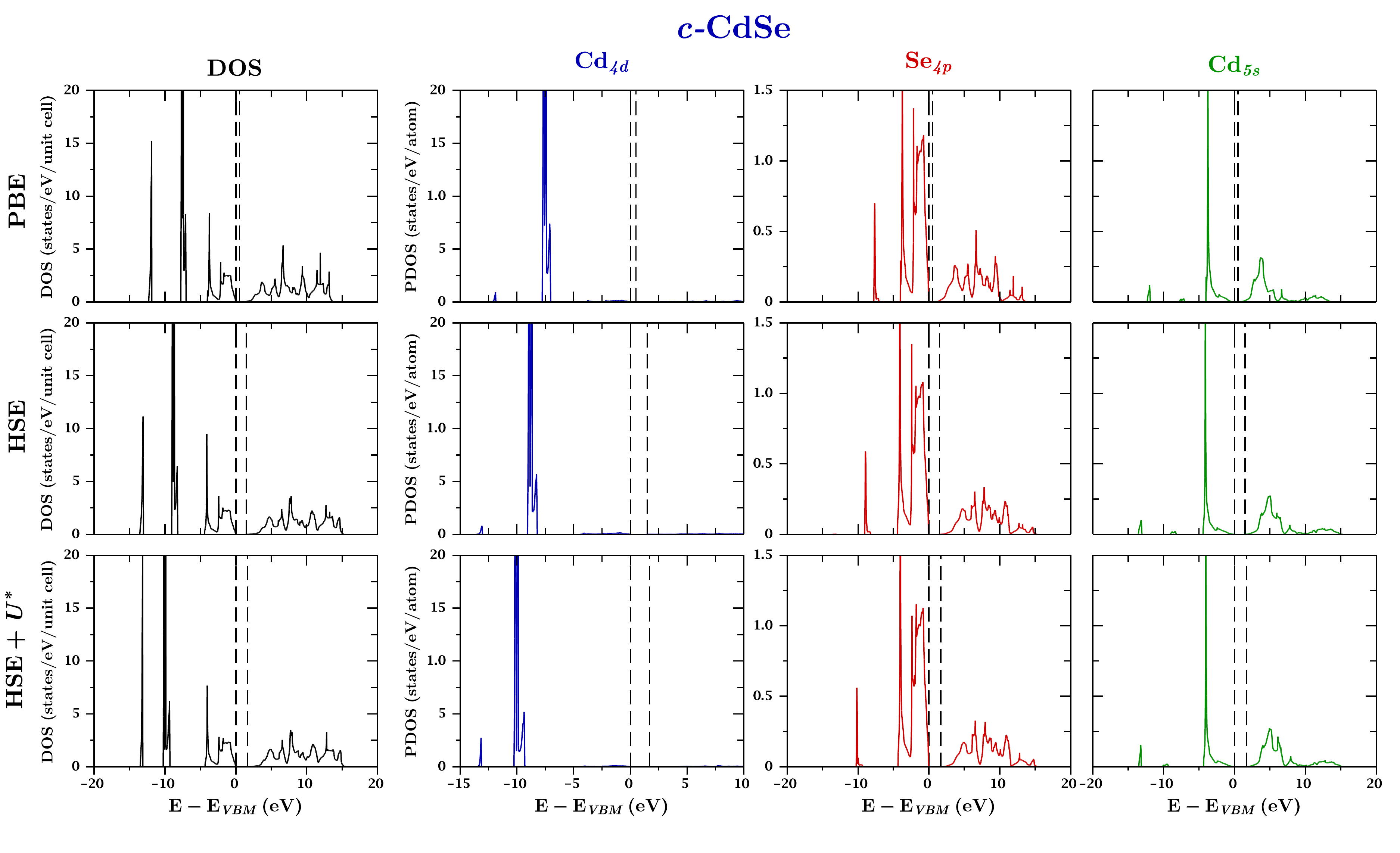}
    }
  \end{center}
  \caption{
          The total and projected density of states of $c$-CdSe.
          }
\end{figure*}

\begin{figure*}
  \begin{center}
    \resizebox{0.94\textwidth}{!}{%
      \includegraphics{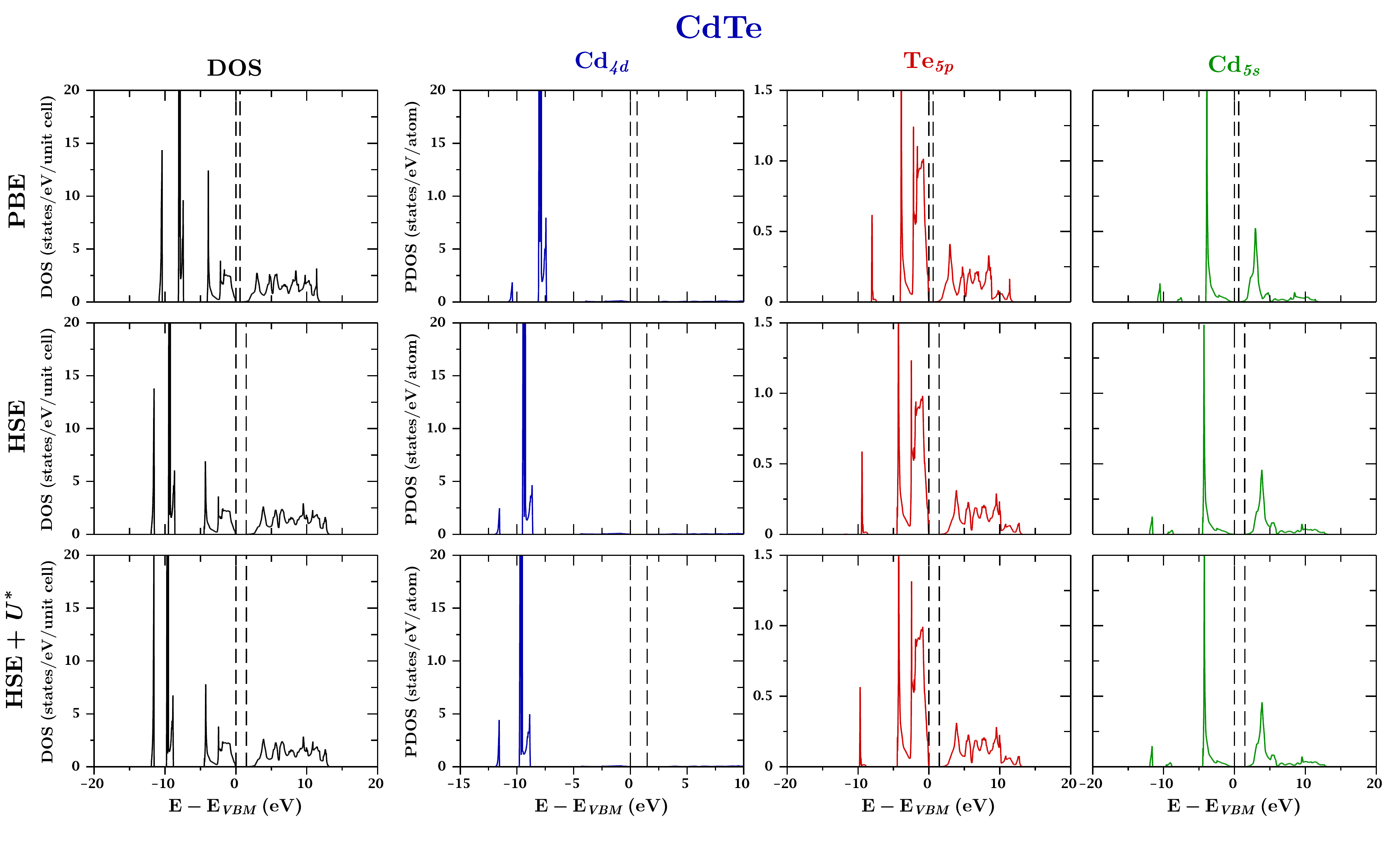}
    }
  \end{center}
  \caption{
          The total and projected density of states of CdTe.
          }
\end{figure*}

\clearpage

\begin{figure}
  \begin{center}
    \resizebox{0.88\textwidth}{!}{%
      \includegraphics{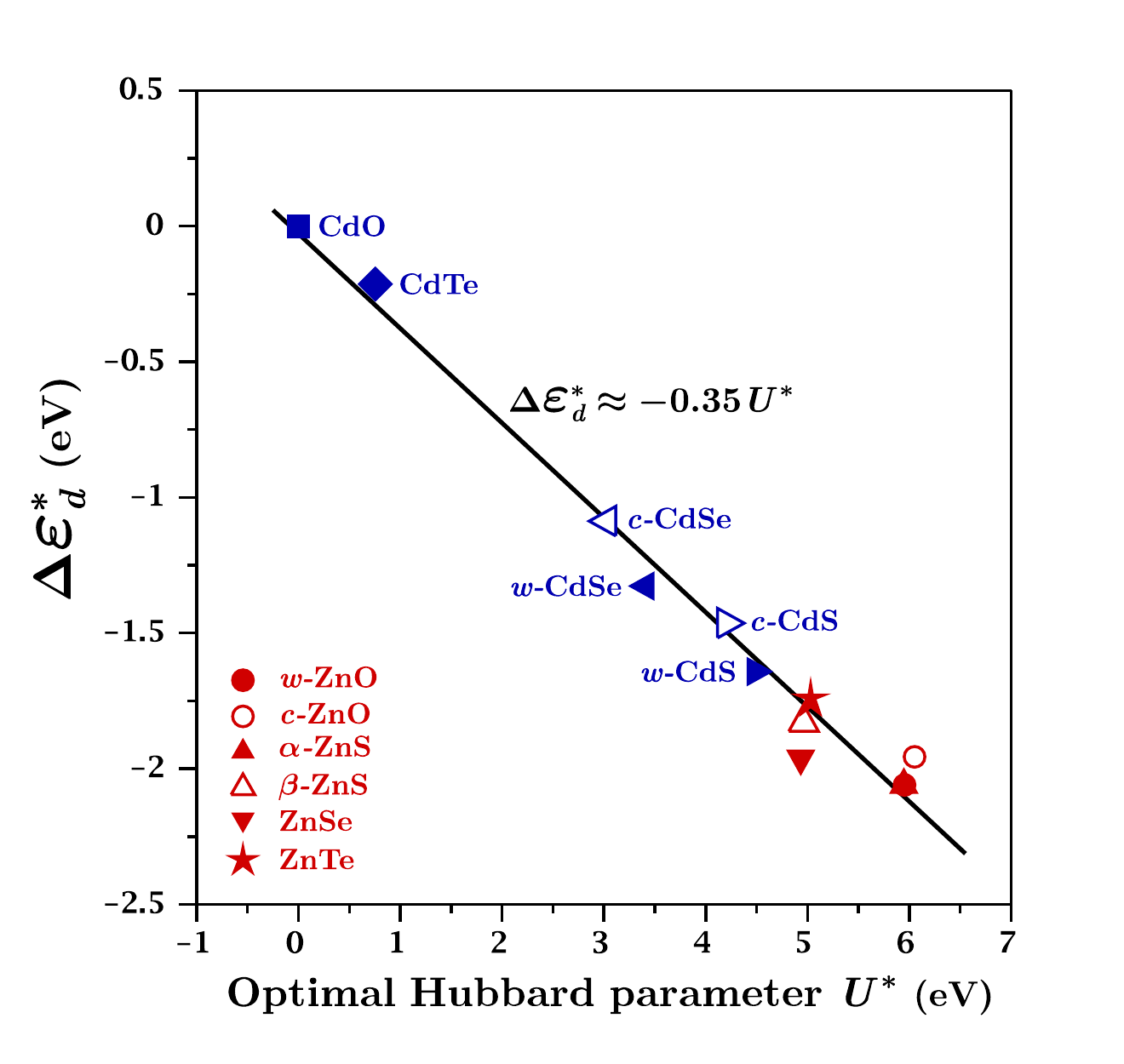}
    }
  \end{center}
  \caption{
          (Color online)
          The difference $\Delta \varepsilon_d^\ast = \varepsilon_d^{{\rm HSE}+U^\ast}- \varepsilon_d^{\rm HSE}$
          as a function of the optimal Hubbard parameter $U^\ast$.
          }
   \label{depsdvsU}
\end{figure}

\end{document}